\documentclass[preprint,5p,times,twocolumn]{elsarticle}
\usepackage{amsmath,amssymb,hyperref,multirow,bm,xcolor}

\journal{Materials Today Communications}

\begin{document}
\begin{frontmatter}
\title{Thermoelectric properties of semiconducting materials with
  parabolic and pudding-mold band structures}
\author[itb,brin]{Jyesta~M.~Adhidewata\corref{cp}}
\ead{jyesta.ma@gmail.com}
\author{Ahmad~R.~T.~Nugraha\corref{cp}}
\ead{ahmad.ridwan.tresna.nugraha@brin.go.id}
\author[brin,unilu]{Eddwi H.~Hasdeo}
\author[rennes]{Patrice Estell\'e} 
\author[itb]{Bobby~E.~Gunara}
\cortext[cp]{Corresponding author}
\address[itb]{Theoretical Physics Research Division,
Faculty of Mathematics and Natural Sciences, Institut Teknologi
Bandung, Bandung 40132, Indonesia}
\address[brin]{Research Center for Physics,
National Research and Innovation Agency (BRIN), 
South Tangerang 15314, Indonesia}
\address[unilu]{Department of Physics and Materials Science,
University of Luxembourg, L-1511 Luxembourg, Luxembourg}
\address[rennes]{Laboratoire de G\'enie Civil
  et G\'enie M\'ecanique (LGCGM), University of Rennes 1, F-35000
  Rennes, France}
\begin{abstract}
By a combination of semi-analytical Boltzmann transport and first-principles calculations, we systematically investigate thermoelectric properties of semiconducting (gapped) materials by varying the degrees of polynomials in their energy dispersion relations, in which either the valence or conduction energy dispersion depends on the wave vector raised to the power of two, four, and six.  Within the relaxation time approximation, we consider various effects such as band gaps, dimensionalities, and dispersion powers to understand the conditions that can give the optimal thermoelectric efficiency or figure of merit ($ZT$).  Our calculations show that the so-called pudding-mold band structure produces larger electrical and thermal conductivities than the parabolic band, but no significant difference is found in the Seebeck coefficients of the pudding-mold and parabolic bands.  Tuning the band gap of the material to an optimum value simultaneously with breaking the band symmetry, the largest $ZT$ is found in a combination of two-contrasting polynomial powers in the dispersion relations of valence and conduction bands.  This band asymmetry also shifts the charge neutrality away from the undoped level and allows optimal $ZT$ to be located at a smaller chemical potential.   We expect this work to trigger high-throughput calculations for screening of potential thermoelectric materials combining various polynomial powers in the energy dispersion relations of semiconductors.  We give preliminary screening results for bulk PtS$_2$ and FeAs$_2$ compared with Si, where we indicate that the former two have better thermoelectric performance than the latter.
\end{abstract}
\end{frontmatter}
\section{Introduction}
Thermoelectric materials can convert a temperature gradient into electricity and vice versa, hence allowing them to have a wide array of potential applications from power generation to refrigeration~\cite{goldsmid10-TEintro}.  A good thermoelectric material or device should produce a high potential difference from a small temperature gradient, with low heat dissipation from ohmic effects, while also thermally insulating to maintain the temperature gradient required for the operation of the device.  In other words, the thermoelectric material ideally should possess a high Seebeck coefficient ($S$) and a high electrical conductivity ($\sigma$) so that the power factor $\mathrm{PF} = S^2 \sigma$ is high enough.  At the same time, the ideal thermoelectric materials should also have a low thermal conductivity $\kappa$.  Therefore, to maximize the efficiency of a thermoelectric material, the dimensionless figure of merit ($ZT$), which is defined as~\cite{goldsmid10-TEintro}
\begin{equation}
    ZT = \frac{S^2 \sigma}{\kappa} T,
    \label{eq:ZT}
\end{equation}
should be as large as possible (at least, usually, $ZT > 1$).

The requirements outlined above pose a well-known thermoelectric trade-off problem~\cite{goldsmid10-TEintro,vining09-TEinconvenient}.  The material with a large Seebeck coefficient tends to have a low carrier concentration, leading to a lower conductivity~\cite{vining09-TEinconvenient,snyder08}.  Therefore, one should properly tune the chemical potential of the thermoelectric material (e.g., by doping) to maximize its power factor~\cite{mahan96-bestTE}.  Meanwhile, the ratio of the electrical conductivity $\sigma$ to the electronic thermal conductivity $\kappa_e$ is usually a constant due to the Wiedemann-Franz law, leading to extensive efforts in obtaining a material with high $\sigma$ but low \emph{lattice} thermal conductivity $\kappa_{ph}$~\cite{zhao2014ultralow, takahashi12} (noting that $\kappa = \kappa_e + \kappa_{ph}$).  

It has often been proposed that materials with non-parabolic band dispersion might overcome these problems~\cite{chen2013}, for example, the quartic "Mexican-hat" band~\cite{wickramaratne15-inse, rudderham20}, the "pudding-mold" band with flat band edge~\cite{kuroki07,kuroki13}, "camel-back" band with multiple valleys~\cite{wang14}, octic ring-shaped bands with a moat~\cite{rudderham21}, or bands with an anisotropic dispersion which are studied generally in \cite{mecholsky14-bandwarping}. The increase in TE performance caused by these non-parabolic band shapes is usually attributed to the effects of the dispersion on the density of states. The presence of a ring-shaped band as in Refs. \cite{wickramaratne15-inse} and \cite{rudderham21} is associated with a singular and discontinuous density of states near the band edge, which was expected to provide a high Seebeck coefficient due to the dependence of the Seebeck coefficient on the derivative of the density of states through the Mott formula. The non-parabolic bands also generally increase the density of states compared to regular bands, thus increasing the conductivity without significantly affecting the Seebeck coefficient, as in the anisotropic band studied in Ref. \cite{chen2013}, or the pudding mold band \cite{kuroki07, kuroki13, kuroki17}.

The "pudding-mold" band, i.e. a band with the electronic energy proportional to the fourth power or higher order of the electronic wave vector, as studied by Usui and Kuroki in Ref. \cite{kuroki17}, is a particularly simple non-parabolic band that shows some of these features, i.e., a larger density of states compared to the parabolic band and the presence of a singularity in the band edge. This simplicity allows us to calculate and compare the various physical parameters of the band to better understand the physical mechanism of any possible thermoelectric performance enhancement, in the hope of finding a way to optimize the TE performance of the material by manipulating its band structure.
As shown in Ref.~\cite{kuroki07}, the combination of a flat portion and a dispersive portion in pudding-mold bands can lead to a high Seebeck coefficient when the chemical potenntial is near to the flat portion due to a large difference in electron velocity (and density of states) above and below the chemical potential. Furthermore, the large electron velocity and large Fermi surface also leads to a higher electron conductivity, which can improve the power factor and the figure of merit. Several other experimental and first-principles studies on materials with the pudding-mold band structures have confirmed their potential uses in thermoelectric applications~\cite{wolverton19,wei20}.

One intriguing direction that remains unexplored for pudding-mold bands is the behavior of a pudding-mold material with \textit{two} bands, in the narrow band gap regime such that the bipolar effect on transport is present. Prior works ~\cite{mahan89-zt,sofo94-optgap,hasdeo19} have shown that there is often an optimal value of band gap which will optimize the value of the power factor or the figure of merit. As such, we think it is worthwhile to expand the previous work concerning this pudding mold type band into two-band systems to determine the band gap dependence of its TE properties and to determine the optimum value of the band gap if any. Furthermore, as the main disadvantage of using narrow gap materials is the presence of bipolar effect from minority carriers to the Seebeck coefficient, \cite{chasmar1959thermoelectric}, we also consider a two-band model in which the dispersion relations of the conduction and the valence bands are of different forms to reduce the bipolar effect through the creation of an imbalance in each carrier transport function.

In this study, we calculate the thermoelectric properties of a
semiconducting material within the two-band model, having a valence
band and a conduction band.  We investigate the effects of various
band shapes and band gaps on its thermoelectric properties.  As recent
advances in thermoelectrics have shown that low-dimensional materials
exhibit good thermoelectric properties due to the quantum confinement
effect \cite{hicks93-qweff, hicks96-TEexp,
  heremans2013thermoelectrics, hung16-quantum}, we also calculate the
corresponding thermoelectric properties for three-dimensional (3D),
two-dimensional (2D), and one-dimensional (1D) cases and compare the
results.  The calculations are performed using the Boltzmann transport
theory along with the relaxation time approximation (RTA).
Furthermore, by first-principles methods, we simulate a couple of
materials with the pudding-mold like band structures, such as Pt$S_2$
and FeAs$_2$, compared with commonly used semiconductors, such as
$Si$, to show their potential for thermoelectric applications.  For
this simulation, the band structure calculations are performed using
the Quantum ESPRESSO package~\cite{QE-2017}, while the thermoelectric
properties are calculated using the BoltzTraP
package~\cite{madsen2006boltztrap,BoltzTraP2}.

This paper is organized as follows.  In Sec.~\ref{sec:th}, we show the
model of the electronic band structures considered in this study, from
parabolic, quartic, to sextic energy dispersion, characterized by the
degrees of polynomials in the dispersion relations.  We also show the
corresponding density of states (DOS) in various dimensions.  We then
describe the semianalytical methods for calculating the thermoelectric
transport coefficients within the RTA.  In Sec.~\ref{sec:res}, we show
the calculated results of the thermoelectric properties for several
combinations of band shapes, such as symmetric bands (valence and
conduction bands having the same degree of polynomials in their energy
dispersion) and asymmetric bands (valence and conduction bands having
different degrees of polynomial in their energy dispersion).  We also
discuss and compare these semianalytical results with the realistic
material simulation.  We finally give the conclusions and perspective
in Sec.~\ref{sec:con}.  Note that to support open science, we provide
Python codes in a Github repository~\cite{githubpuddingTE} for the
readers to reproduce our results.

\section{Model and methods}
\label{sec:th}

For the semianalytical calculations of the thermoelectric transport
coefficients ($S$, $\sigma$, and $\kappa_e$), we model dispersion
relations for the conduction and valence bands by
\begin{align}
  E_c(\mathbf{k}) &= (A_m \vert \mathbf{k}\vert^m + \Delta),
  \label{eq:cdis}
\end{align}
and
\begin{align}
  E_v(\mathbf{k}) &= -(A_n \vert \mathbf{k}\vert^n + \Delta),
  \label{eq:vdis}
\end{align}
respectively, where $m$ and $n$ are even integers (2, 4, 6), $A_m$
($A_n$) is the normalization constant that depends on $m$ ($n$),
set such that the band width is unchanged for different values of $m$ or $n$,
$\Delta$ is half of the band gap $E_g$, and $\mathbf{k}$ is the
electronic wave vector.
Following Usui and Kuroki~\cite{kuroki17}, we take the normalization
constant $A_n$ to be
\begin{equation}
    A_n = \left(\frac{a}{\sqrt{d}\pi}\right)^n E_0.
    \label{sym:norm}
\end{equation}
With this normalization, for a $d$-dimensional cubic unit cell with lattice constant $a$, at the edge of the first Brillouin zone for any value of $n$, the bands will converge at the same band width $E_0$. In our calculations, we set the band edge to a reasonable value of $2$ eV.  We show in \ref{append:bandwidth} that the results are also actually quite insensitive to different values of $E_0$.
Using dispersion relations of Eqs.~\eqref{eq:cdis} and~\eqref{eq:vdis}, we can calculate the DOS, denoted by
$g(\epsilon)$, as follows:
\begin{equation}
  g(\epsilon) = C_d a^{d-3} \int \delta(\epsilon - E(\mathbf{k})) k^{d-1} dk,
  \label{sym:dimcont}
\end{equation}
where $\epsilon$ is the electronic energy and $C_d$ is a constant that
depends on dimension: $C_1 = 1/\pi $, $C_2 = 1/2\pi$, and
$C_3 = 1/2\pi^2$.  For the conduction band we have
\begin{equation}
  g(\epsilon) = 
  \frac{C_d}{mA_m a^{3-d}} \left(\frac{\epsilon - \Delta}{A_m} \right)^{(d-m)/m},
  \label{eq:DOSc}
\end{equation}
while for the valence band we have
\begin{equation}
  g(\epsilon)=
  \frac{C_d}{nA_n a^{3-d}} \left(\frac{-\epsilon - \Delta}{A_n} \right)^{(d-n)/n}.
\end{equation}

\begin{figure}[tb]
  \centering\includegraphics[width=85mm,clip]{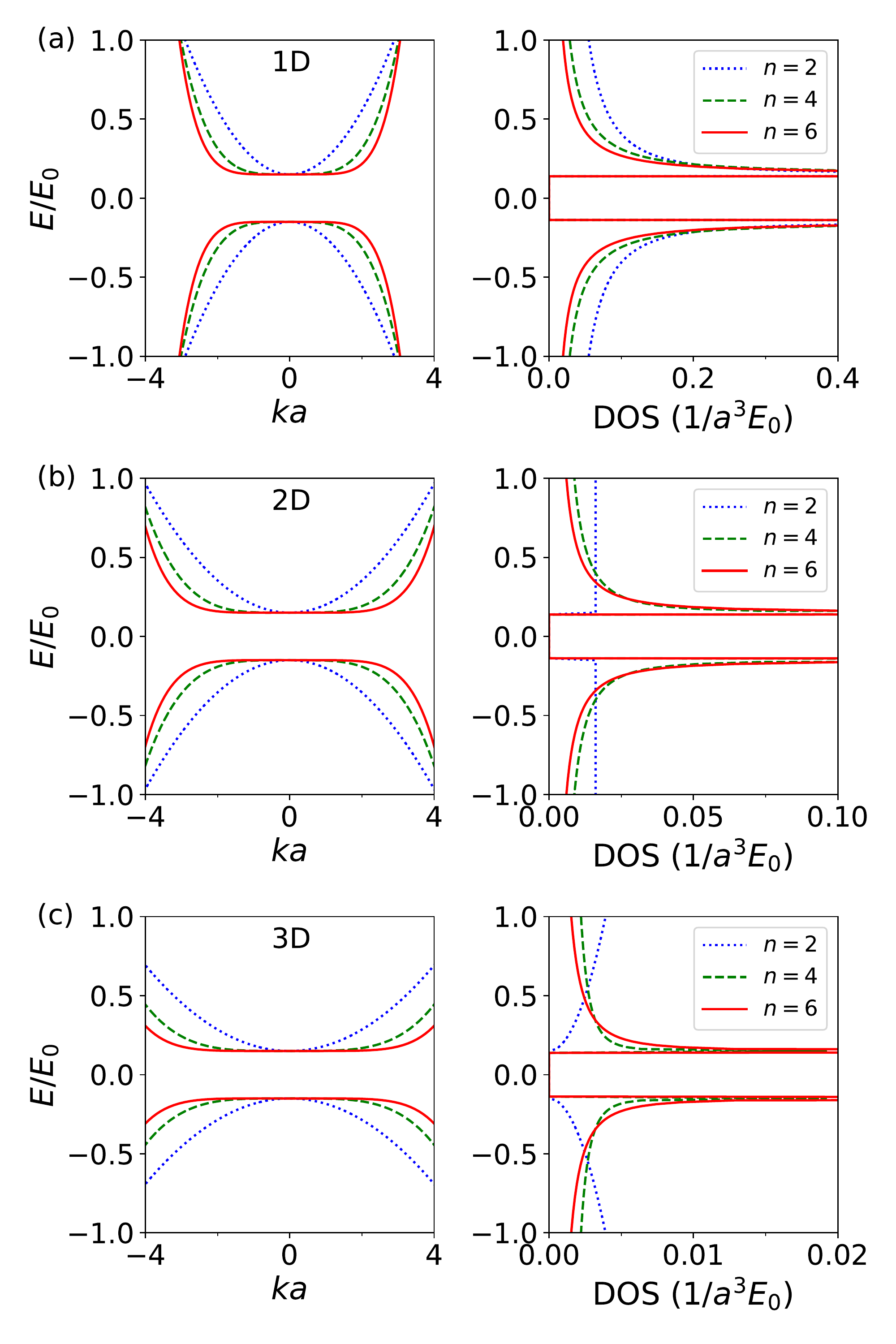}
  \caption{The energy dispersion and density of states for different
    values of $m = n$ in (a) 1D, (b) 2D, and 3D cases. In this plot we
    deliberately choose $\Delta = 0.15 E_0$}
    \label{figTEDOS}
\end{figure}

Schematic plots of the energy dispersion and its corresponding DOS for different dimensions of the material in the case of $m = n$ can be seen in Figs.~\ref{figTEDOS}(a)--(c).  We see that for the pudding-mold band, the DOS diverges near the band edge, which is more similar to the density of states of a one-dimensional parabolic band as shown in Fig.~\ref{figTEDOS}(1) than that of the usual 2D or 3D parabolic band. Indeed, from Eq.\eqref{eq:DOSc} we see that the density of states of the two-dimensional quartic band and the three-dimensional sextic band has the same form as a 1D parabolic band. However, the effect of the pudding mold band is more than a simple dimensional reduction, as from Eq. \eqref{eq:DOSc} it is clear that other values of $d$ and $n$ does not correspond to a lower-dimensional band, and this is disregarding the effect of density of states on the electron velocity. Assuming the materials to be isotropic in energy, we can also obtain the longitudinal velocity $v$ of electrons in a certain direction, $v^2 = v_g^2/d$, where $v_g$ is the group velocity.  Then, since $v_g = \hbar^{-1} dE/dk$, we obtain
\begin{equation}
  v = \pm \frac{nA_n}{\hbar \sqrt{d}} k^{n-1}.
\end{equation}

Using Boltzmann transport theory with the relaxation time
approximation, the Seebeck coefficient, the electrical conductivity,
and the electronic thermal conductivity can be expressed as
\cite{ashcroft, hasdeo19}
\begin{align}
  S &= \frac{1}{eT} \frac{\mathcal{L}_1}{\mathcal{L}_0}, \\
  \sigma &= e^2 \mathcal{L}_0, 
\end{align}
and
\begin{align}
  \kappa_e = \frac{1}{T} \left(\mathcal{L}_2 -
             \frac{\mathcal{L}_1^2}{\mathcal{L}_0} \right),
\end{align}
respectively, where $T$ is the absolute temperature, and
$\mathcal{L}_i$ is the ``thermoelectric integral'':
\begin{equation}
  \mathcal{L}_i = \int \tau v^2 g(\epsilon) 
  \left(-\frac{\partial f}{\partial \epsilon}\right) 
  (\epsilon - \mu)^i d\epsilon ~ ,
  \label{eq:therminteg}
\end{equation}
where $\tau$ is the relaxation time.

Since our model materials have two separate bands, we can write the
total thermoelectric integral as a sum of conduction and valence band
components given by
\begin{align}
  \mathcal{L}_{c,i} = \int_{\Delta}^{\infty} \tau v^2 g(\epsilon)
                      \left(-\frac{\partial f}{\partial
                      \epsilon}\right)
                      (\epsilon - \mu)^i d\epsilon
\end{align}
and
\begin{align}
  \mathcal{L}_{v,i} = \int_{-\infty}^{-\Delta} \tau v^2 g(\epsilon) 
                      \left(-\frac{\partial f}{\partial
                      \epsilon}\right)
                      (\epsilon - \mu)^i d\epsilon, 
\end{align}
respectively, so that the thermoelectric properties of the materials
can be expressed as a combination of the contribution from each band:
\begin{align}
  S &= \frac{S_c \sigma_c + S_v \sigma_v}{\sigma_c + \sigma_v}, \\
  \sigma &= e^2 (\mathcal{L}_{c,0} + \mathcal{L}_{v,0}) 
           = \sigma_c + \sigma_v,
\end{align}
and
\begin{align}
    \kappa_e = \frac{\sigma_c \sigma_v}{\sigma_c + \sigma_v}(S_c - S_v)^2
             + (\kappa_{e,c} + \kappa_{e,v}).
    \label{eq:eachbandTE}
\end{align}

By substituting $v$ and $g(\epsilon)$ to the thermoelectric integrals
$\mathcal{L}_{c,i}, \mathcal{L}_{v,i}$, we have for the conduction
band,
\begin{align}
  \mathcal{L}_{c,i} =& \frac{C_d m A_m^{(2-d)/m}}{da^{3-d}\hbar^2} \times \notag\\
                     &\int_{\Delta}^{\infty} \tau (\epsilon -
                       \Delta)^{1+(d-2)/m}(\epsilon-\mu)^i 
                       \left(\frac{\partial f_0}{\partial \epsilon}
                       \right) 
                       d\epsilon,
    \label{eq:Lic}
\end{align}
and similarly for the valence band,
\begin{align}
  \mathcal{L}_{v,i} =& \frac{C_d n A_n^{(2-d)/n}}{da^{3-d}\hbar^2}
                       \times \notag\\
                     &\int_{-\infty}^{\Delta} \tau 
                       (-\epsilon -
                       \Delta)^{1+(d-2)/n}(\epsilon-\mu)^i 
                       \left(\frac{\partial f_0}{\partial \epsilon}
                       \right) 
                       d\epsilon.
    \label{eq:Liv}
\end{align}

To proceed with our calculation, we need to determine the dependence
of $\tau$ on $\epsilon$.  This dependence is determined by the scattering mechanism of the electron. Generally in TE transport the dominant scattering is caused by electron-impurity scattering or electron-phonon interaction, which results in either a constant relaxation time in the case of electron-impurity scattering, or a $\tau$ that is inversely proportional to DOS in case of electron-phonon interaction~\cite{hung17inse, rudderham20}.  The constant relaxation time approximation (CRTA) can adequately explain experimental results for some materials~\cite{schiedemantel03, singh10}, including the case of pudding-mold band structures~\cite{kuroki13}.  Therefore, we decide to adopt the CRTA in this work. To check the validity of our results with this approximation we perform similar calculations for $\tau = 1/\mathrm{DOS}$, or the DOS dependent relaxation time approximation (DRTA), as given in \ref{append:drta}. 

\begin{figure*}[t!]
  \centering\includegraphics[width=16cm,clip]{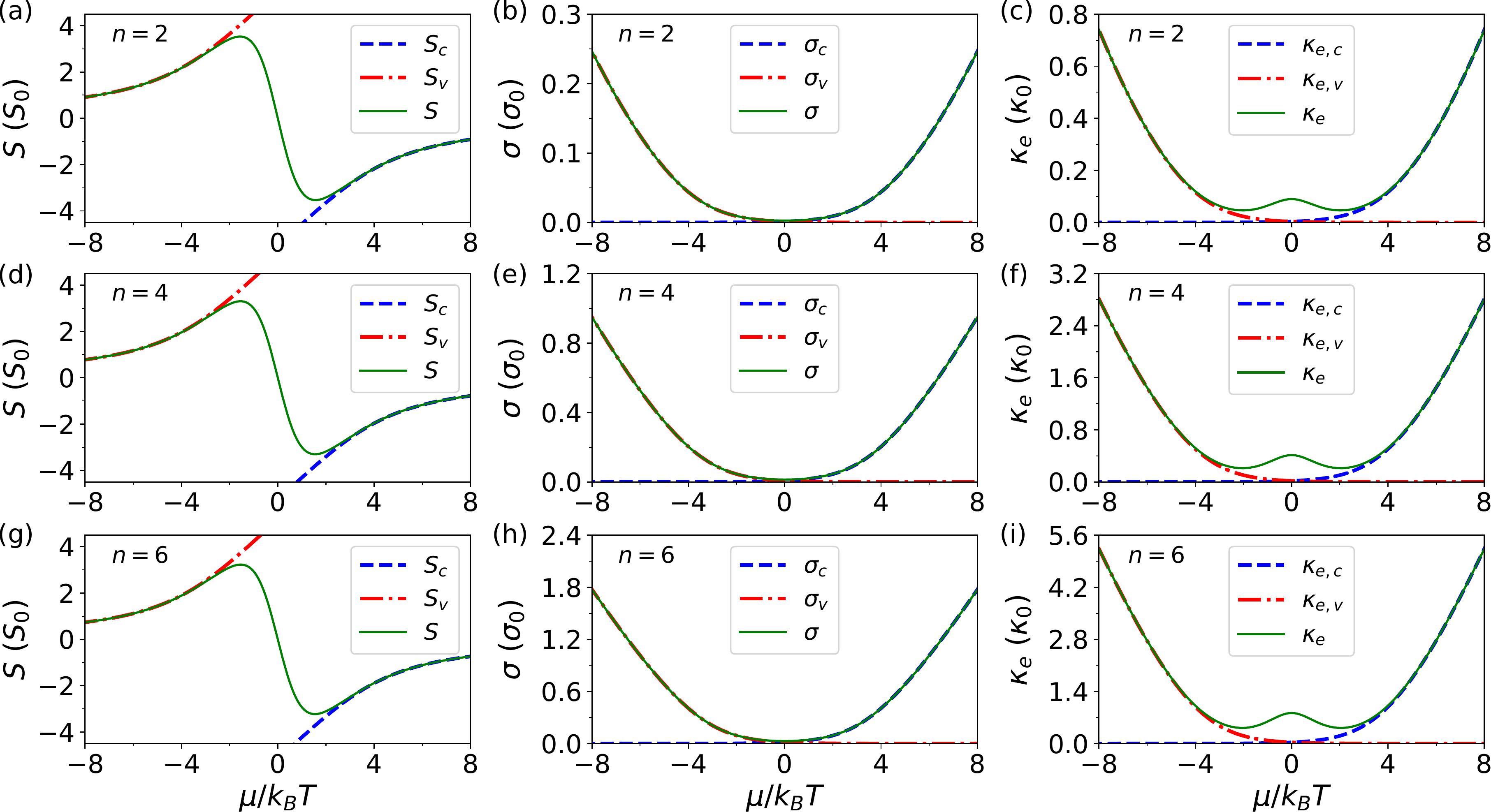}
  \caption{Seebeck coefficient ($S$), electrical conductivity
    ($\sigma$), and electronic thermal conductivity ($\kappa_e$)
    considering symmetric ($m = n$) conduction and valence bands with
    $n = 2$ in (a)--(c), $n = 4$ in (d)--(f), and $n = 6$ in (g)--(i).
    The results for $S$, $\sigma$, and $\kappa_e$ are expressed in the
    units of $S_0$, $\sigma_0$, and $\kappa_0$ respectively.  The band
    gap is set to be $E_g = 6 k_B T$, which corresponds to
    $\Delta = 3k_B T$.}
  \label{fig2}
\end{figure*}

Let $(\epsilon - \mu)/k_B T = x$, $\tilde{\Delta} = \Delta/k_B T$, and
$\eta = \mu/k_B T$.  We can rewrite the thermoelectric integrals for
the conduction band:
\begin{equation}
  \mathcal{L}_{c,i} = 
  \frac{C_d \tau m (k_B T)^{i+1} }{a^{3-d}d\hbar^2} 
  \left(\frac{A_m}{k_B T}\right)^{(2-d)/m} 
  \mathcal{G}_{c,i}(\eta, \tilde{\Delta}),
\end{equation}
with
\begin{equation}
  \mathcal{G}_{c,i}(\eta, \tilde{\Delta})
  = \int_{\tilde{\Delta}-\eta}^{\infty}
  \frac{x^i e^x (x+\eta-\tilde{\Delta})^{1+(d-2)/n}}{(e^x+1)^2} dx,
  \label{eq:Gc}
\end{equation}
and for the valence band:
\begin{equation}
  \mathcal{L}_{v,i} = 
  \frac{C_d \tau n (k_B T)^{i+1} }{a^{3-d}d\hbar^2}
  \left(\frac{A_n}{k_B T}\right)^{(2-d)/n} 
  \mathcal{G}_{v,i}(\eta, \tilde{\Delta}),
\end{equation}
with
\begin{equation}
  \mathcal{G}_{v,i}(\eta, \tilde{\Delta}) =
  \int_{\tilde{-\infty}}^{-\eta-\tilde{\Delta}}
  \frac{x^i e^x (-x-\eta-\tilde{\Delta})^{1+(d-2)/n}}{(e^x+1)^2} dx.
    \label{eq:Gv}
\end{equation}
Now we can express the thermoelectric transport coefficients in terms
of $\mathcal{G}_{c,i}, \mathcal{G}_{v,i}$ as follows:
\begin{align}
  S_c =& S_0 \frac{\mathcal{G}_{c,1}
         (\eta, \tilde{\Delta})}{\mathcal{G}_{c,0}
         (\eta, \tilde{\Delta})}, \label{eq:Sfinal}\\
  \sigma_c =& \frac{C_d na^{d-2}}{d} 
              \left(\frac{A_n}{k_B T}\right)^{(2-d)/n} 
              \sigma_{0} \mathcal{G}_{c,0}
              (\eta, \tilde{\Delta}), \label{eq:Sigmafinal}
\end{align}
and
\begin{align}
  \kappa_{e,c} =& \frac{C_d na^{d-2}}{d} 
                  \left(\frac{A_n}{k_B T}\right)^{(2-d)/n} 
                  \kappa_{0} \notag\\
                &\times \left(\mathcal{G}_{c,2}(\eta, \tilde{\Delta}) - 
                  \frac{\mathcal{G}_{c,1}
                  (\eta, \tilde{\Delta})^2}{\mathcal{G}_{c,0}
                  (\eta, \tilde{\Delta})}\right). \label{eq:kappafinal}
\end{align}
The normalization constants $S_0$, $\sigma_0$, and $\kappa_0$ are
given by
\begin{align}
    S_0  &= \frac{k_B}{e} = 86.17~ \mathrm{\mu V/K}, \\
    \sigma_0 &= \frac{e^2 \tau k_B T}{a\hbar^2},
\end{align}
and
\begin{align}
    \kappa_0 &= \frac{\tau k^3_B T^2}{a\hbar^2}.
    \label{eq:kappanorm}
\end{align}
As an estimate for the value of $\sigma_0$ and $\kappa_0$, if we take
$\tau = 10^{-14}~\mathrm{s}$, $T = 350~\mathrm{K}$, and
$a = 10^{-10}~\mathrm{m}$, we obtain
$\sigma_0 = 1.12 \times 10^6~\mathrm{S/m}$ and
$\kappa_0 = 2.89~ \mathrm{W/m K}$.  Similar relations for the valence
band are obtained by swapping
$\mathcal{G}_{c,i}(\eta, \tilde{\Delta})$ to
$\mathcal{G}_{v,i}(\eta, \tilde{\Delta})$ in the previous equations.
We will use these normalization constants in all calculated results in
Secs.~\ref{subsec:symm}--\ref{subsec:dim}.   

\section{Results and Discussion}
\label{sec:res}

Now we can calculate the thermoelectric properties of the model
material using Eqs.~\eqref{eq:Sfinal}--\eqref{eq:kappafinal}, where
the integrals $\mathcal{G}_{(c,v),i}$ are calculated numerically using
the SciPy package in Python~\cite{2020SciPy-NMeth}.  We start with 3D
systems in Secs.~\ref{subsec:symm} and~\ref{subsec:asymm}.  Then, we
explore the cases of lower dimensions (2D and 1D systems) in
Sec.~\ref{subsec:dim}.  To get insight into the real systems as
thermoelectric materials, we give the preliminary screening of a
couple of semiconductors in Secs.~\ref{subsec:real}

\subsection{Symmetric two-band systems}
\label{subsec:symm}
First, we will show the thermoelectric properties for symmetric
conduction and valence bands, i.e. $m = n$.  In all the calculations
below, we take a reference temperature $T_0 = 300 ~\mathrm{K}$ such
that $k_B T_0 \approx 0.026~\mathrm{eV}$ and the band width
$E_0 = 2~\mathrm{eV}$, corresponding the reduced variable of thermal
energy $k_B T/E_0 = 0.013$.  We plot the Seebeck coefficient,
electrical conductivity, and electronic thermal conductivity versus
the chemical potential for the parabolic band ($n = 2$) and
pudding-mold bands ($n = 4$ and $n = 6$) in Figs.~\ref{fig2}(a)--(i)
using $\Delta = 3k_B T$ ($E_g = 6k_B T$), in which we express the
thermoelectric properties in terms of the normalization constants
$S_0$, $\sigma_0$, and $\kappa_0$.

\begin{figure*}[t!]
  \centering\includegraphics[width=16cm,clip]{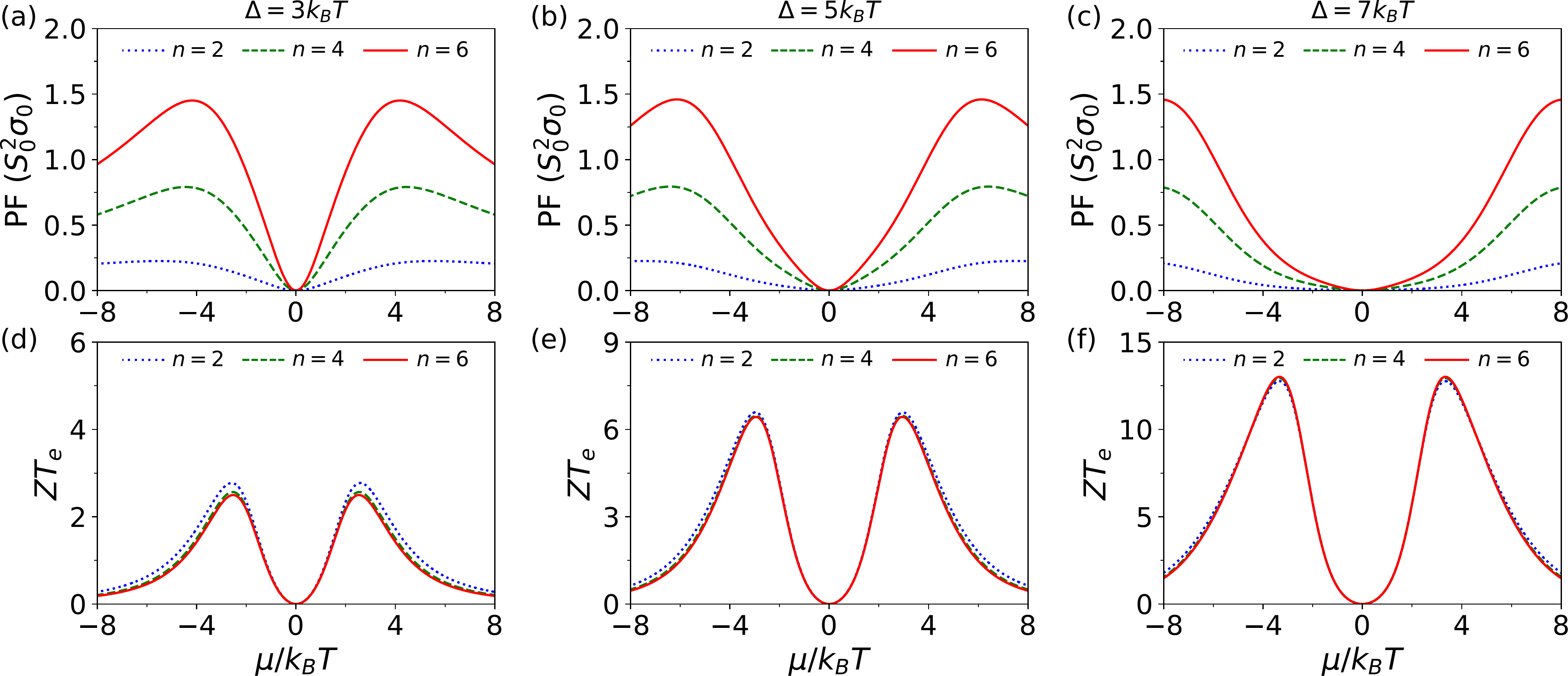}
  \caption{Power factor (PF) and electronic figure of merit ($ZT_e$)
    considering symmetric conduction and valence bands with
    $n = 2, 4$, and $6$.  Panels (a-c) show the results for PF, while
    panels (d-g) for $ZT_e$.  The band gaps are varied corresponding
    to $\Delta = 3k_B T$ ($E_g = 6 k_B T$) in (a) and (d),
    $\Delta = 5k_B T$ ($E_g = 10 k_B T$) in (b) and (e), and
    $\Delta = 7k_B T$ ($E_g = 14 k_B T$) in (c) and (f).}
  \label{fig3}
\end{figure*}

From Figs.~\ref{fig2}(b)--(c), \ref{fig2}(e)--(f),
and~\ref{fig2}(h)--(i) we see that the values of conductivities at the
same chemical potential are highly dependent on the band shapes ($n$
values), while the Seebeck coefficients are almost unaffected by the
different band shapes.  The dependence of electrical conductivity on
the band shape comes from the density of states and the mean electron
velocity due to the presence of the $g(\epsilon) v^2 (\epsilon)$
factor in the thermoelectric integral [Eq.~\eqref{eq:therminteg}].
When the energy $\epsilon$ is low, the velocity in pudding-mold band
is lower than the velocity for parabolic band, but the density of
states for pudding-mold band is higher (and diverges as $\epsilon$
nears the band edge).  The interplay of these factors causes higher
total conductivity for pudding-mold band.  When the energy is high,
the density of states for pudding-mold band is lower than the
parabolic band, while the velocity for pudding mold band is higher
than the parabolic band.  However, since it is the square of electron
speed that shows up in the integral, the combined result is that the
factor $g(\epsilon) v^2(\epsilon)$ is always higher for a pudding-mold
band in the 3D case.  Note that for the 1D case, as will be shown
later in Sec.~\ref{subsec:dim}, the conductivity for the parabolic
band is higher than for the pudding mold bands when the chemical
potential is near the band gap since the density of states for the
parabolic band also diverges near the band gap, unlike the 2D and 3D
case.

We can interpret the thermoelectric integral $\mathcal{L}_0$ as a
quantity characterizing the transport strength due to drift velocity
caused by an external electric field, while $\mathcal{L}_1$ is related
to electron diffusion due to the thermal effects.  A large Seebeck
coefficient can be achieved if the effects of electron drift are not
large enough to neutralize the potential difference caused by the
thermal diffusion.  However, both the electron drift and electron
diffusion have a similar dependence on the
$g(\epsilon) v^2 (\epsilon)$ factor.  Therefore, changing the band
shape has a negligible effect on the Seebeck coefficient as shown in
Figs.~\ref{fig2}(a), \ref{fig2}(d), and~\ref{fig2}(g).  We can also
justify this behavior from the Seebeck coefficient formula,
\begin{equation}
  S = \frac{S_c \sigma_c + S_v \sigma_v}{S_c + S_v} 
  = S_0 \frac{({G}_{c,1}(\eta, \tilde{\Delta})
    +{G}_{v,1}(\eta, \tilde{\Delta}))}{{G}_{v,0}
    (\eta, \tilde{\Delta})+{G}_{v,0}(\eta, \tilde{\Delta})} ~ ,
\end{equation}
in which from Eqs.~\eqref{eq:Gc} and \eqref{eq:Gv} we see that
integrals $\mathcal{G}_i$ only depend slightly on the band shape
through $(x+\eta - \Delta)^{1+1/n}$.  As a result, the Seebeck
coefficient is almost independent of the dispersion power ($n$), while
from equations \ref{eq:Sigmafinal}, \ref{eq:kappafinal} we see that
the electrical and thermal conductivities of electrons depend almost
linearly on $n$.  These results are in agreement with the previous
work from Usui and Kuroki \cite{kuroki17}, in which they considered
only the one-band model without band-gap dependence.

\begin{figure}[t]
  \centering \includegraphics[width=85mm,clip]{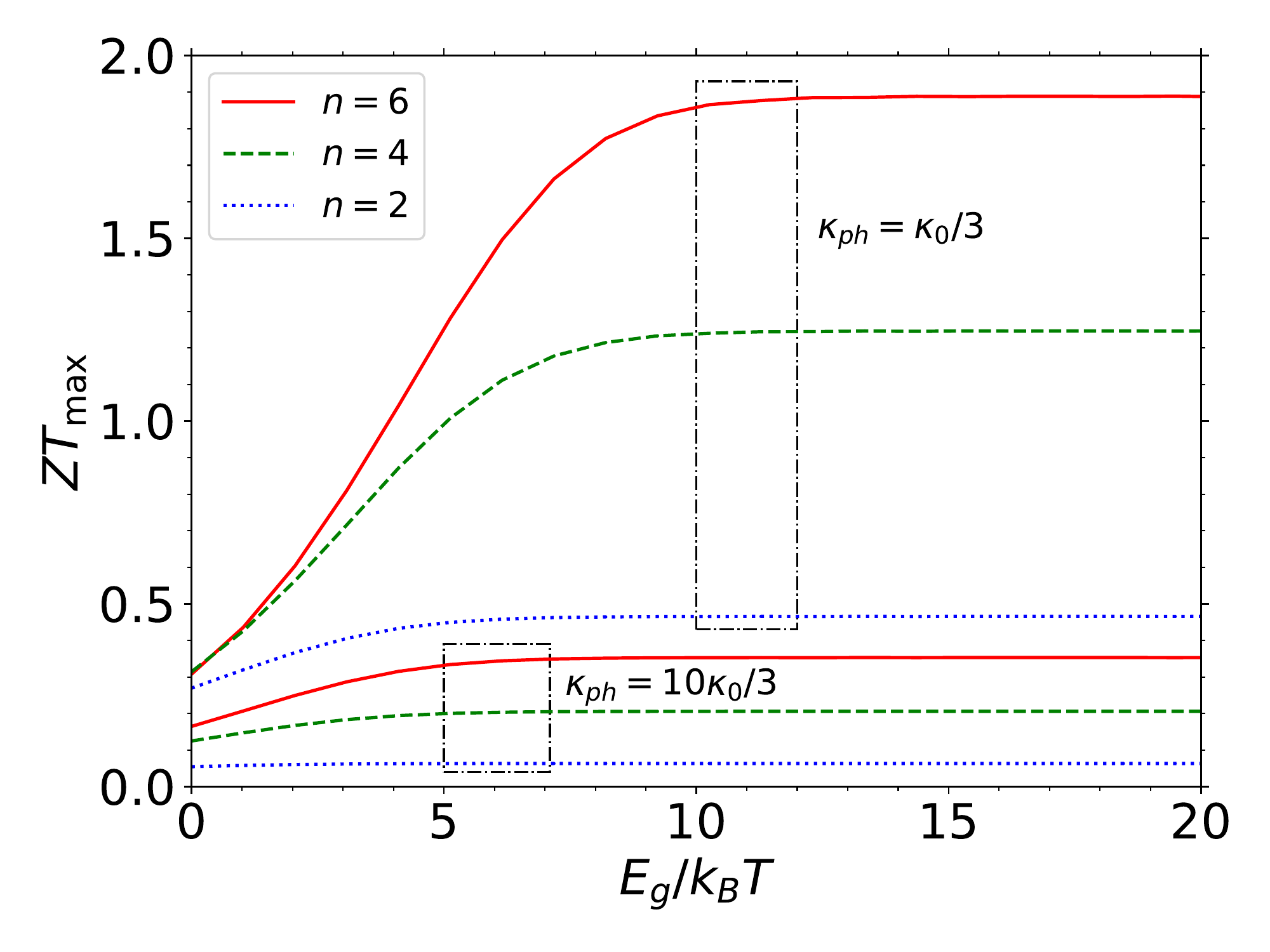}
  \caption{Maximum figure for merit $ZT_{\textrm{max}}$ (at the
    optimum chemical potential) as a function of band gap for several
    cases of symmetric bands with two different parameters of phonon
    thermal conductivity: $\kappa_{ph} = \kappa_0/3$ and
    $\kappa_{ph} = 10 \kappa_0/3$.  Dash-dotted boxes serve as a guide
    for eyes to distinguish the calculated curves obtained from the
    different $\kappa_{ph}$ parameters.}
  \label{fig4}
\end{figure}

Since the power factor of a material is given by
$\mathrm{PF} = S^2 \sigma$, we see that a higher dispersion power $n$
will also lead to a higher $\mathrm{PF}$ due to the increased
conductivity, as can be seen in Figs.~\ref{fig3}(a)--(c).  On the
other hand, since the figure of merit is proportional to
$\sigma/\kappa$ and both conductivities are linear to $n$ when
$\kappa_{ph} = 0$, its dependence on band shape is only caused by the
slightly different Seebeck coefficient.  As a result, increasing the
band shape slightly decreases the figure of merit in the most ideal
case due to the decreased Seebeck coefficient.  The most ideal figure
of merit, which correspond to the so-called electronic figure of merit
($ZT_e$) shown in Figs.~\ref{fig3}(d)--(f), only holds when the phonon
thermal conductivity is zero.  We will consider the non-ideal case
with nonzero $\kappa_{ph}$ shortly.  

One possible way of increasing the Seebeck coefficient is by increasing the band gap. With a narrow band gap, there will be more bipolar conduction such that the Seebeck coefficient from the conduction band and the valence band will cancel each other.  However, a large band gap will also lower the electron conductivity of a material since the number of available charge carriers is lower.  This condition might not give a significant improvement of the power factor of the material.  To clearly see the effects of different band gaps, we calculate the power factors and electronic figures of merit for $\Delta = 3k_B T$~[Figs.~\ref{fig3}(a) and~\ref{fig3}(d)], $\Delta = 5k_B T$~[Figs.~\ref{fig3}(b) and~\ref{fig3}(e)], and $\Delta = 7k_B T$~[Figs.~\ref{fig3}(c) and~\ref{fig3}(f)].

We see that the maximum possible value of the power factor is not affected by the increase in the band gap.  Apparently, the increase in the Seebeck coefficient is perfectly counteracted by the decrease in electrical conductivity. This is because as the band gap increases we can suppress the bipolar effect on the Seebeck coefficient by increasing the doping of our material to optimize this maximum power factor. As a result, after passing a critical point where the effects of the opposing band become small enough, increasing the band gap no longer affects the power factor since we are effectively dealing with only a single band.

On the other hand, for the ideal figure of merit the decrease in
electrical conductivity and thermal conductivity cancels each
other. As such, the peak for the figure of merit when
$\kappa_{ph} = 0$ is identical to the peak of the Seebeck coefficient,
which is located near the middle of the band gap, where the effects of
both band on the Seebeck coefficient is never negligible. This means
increasing the band gap will still increase the figure of merit
provided that $\sigma/\kappa$ remains constant. When
$\kappa_{ph} \neq 0$, as we will see later, $\sigma/\kappa$ will no
longer be constant and similarly to the power factor, past a certain
point increasing the band gap will have no effect on the figure of
merit.

In a real material, the phonon thermal conductivity is not zero and
generally comparable to or larger than the electron thermal
conductivity. As an estimate for the behaviour of figure of merit when
the phonon thermal conductivity is not zero, we calculate the figure
of merit with the phonon thermal conductivity as a constant multiple
of $\kappa_0$, $\kappa_{ph} = \alpha \kappa_0$.  References show that
the values of $\kappa_{ph}$ for materials with pudding-mold band
structures such as $\mathrm{BaPdS}_2$ are in the range of
$1$--$10~\mathrm{W/m.K}$~\cite{wolverton19, wei20}, or, since
$\kappa_0 \approx 3~\mathrm{W/m.K}$, the values of $\kappa_{ph}$ are
around $1/3$ to $10/3$ times $\kappa_0$.  Using this consideration,
in Fig.~\ref{fig7} we plot the peak value of figure of merit
$ZT_{\textrm{max}}$ at an optimized chemical potential $\mu$ (for a
certain band shape and band gap) as a function of band gap.

\begin{figure*}[t!]
  \centering\includegraphics[width=16cm,clip]{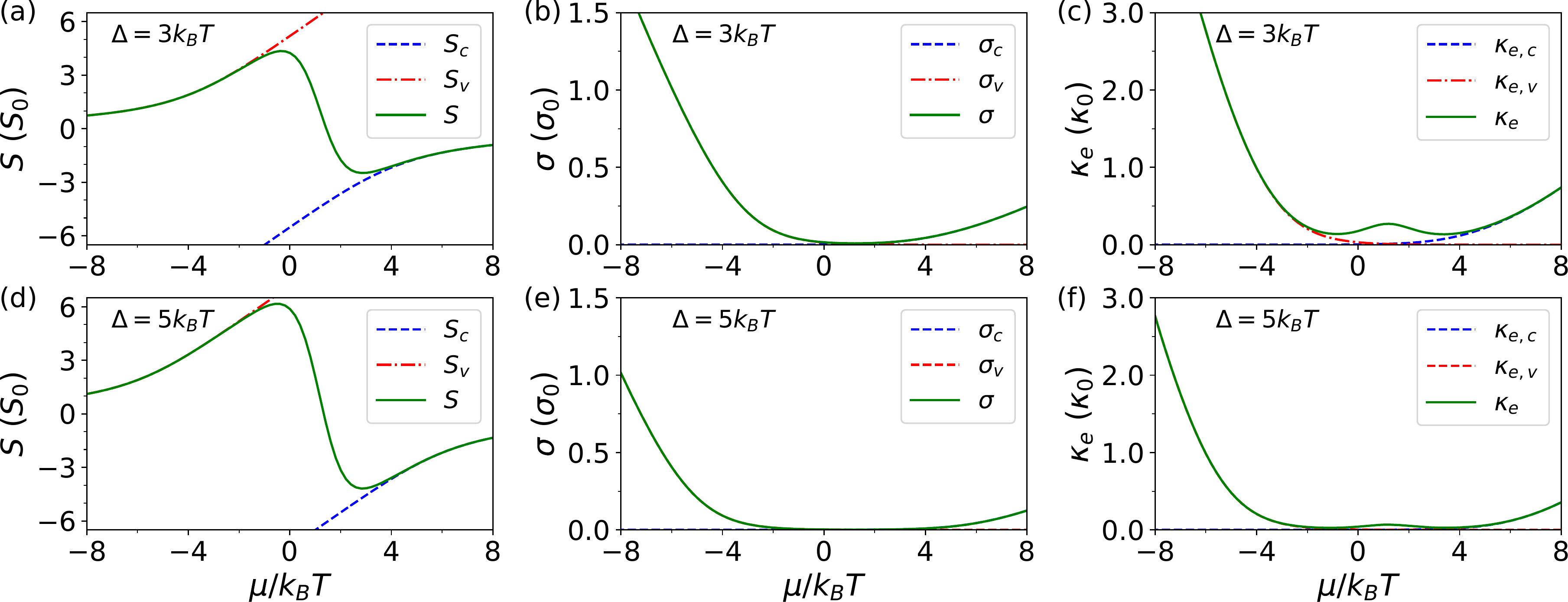}
  \caption{Seebeck coefficient ($S$), electrical conductivity
    ($\sigma$), and electronic thermal conductivity ($\kappa_e$)
    considering asymmetric conduction and valence bands with $m=2$ and
    $n=6$, respectively.  The results for $S$, $\sigma$, and
    $\kappa_e$ are expressed in the units of $S_0$, $\sigma_0$, and
    $\kappa_0$ respectively.  The band gaps are varied corresponding
    to $\Delta = 3k_B T$ ($E_g = 6 k_B T$) in (a)--(c) and
    $\Delta = 5k_B T$ ($E_g = 10 k_B T$) in (d)--f).}
  \label{fig5}
\end{figure*}

From Fig.~\ref{fig4} we see that the introduction of a
nonzero $\kappa_{ph}$ creates a significant difference in the figure
of merit of different band shapes, unlike for the ideal case. We can
see this from the definition of figure of merit,
\begin{equation}
  ZT = \frac{S^2 \sigma}{\kappa_e + \kappa_{ph}}T
  \propto \frac{S^2_0 \sigma_0}{\kappa_0 + \kappa_{ph}/n}T ~ .
\end{equation}
Thus, increasing $n$ brings the value of $\sigma/\kappa$ closer to the
ideal value.  This effect is more noticeable for larger band gap
because as the band gap is increased, $\kappa_{e}$ becomes small and
$\kappa_{ph}$ is the dominant contributor to the total thermal
conductivity.  We also see that in accordance to the previous result,
increasing the band gap will increase the figure of merit, but, when
$\kappa_{ph}$ is nonzero, there will be an asymptotic limit to the
value of figure of merit from increasing the band gap only. This is
because when the band gap is increased, $\kappa_{ph}$ become much
larger compared to $\kappa_e$ such that the figure of merit is
effectively proportional to $\sigma$, i.e.
$ZT \approx S^2 \sigma/\kappa_{ph}$.  Similar to the power factor, the
optimal chemical potential at which $ZT_{\mathrm{max}}$ is pushed
nearer to the band edge, where the effects from the opposing band is
negligible, and thus any further increase in the band gap will not
affect the figure of merit.  The value of $\Delta$ for which the limit
is achieved is dependent on $\kappa_{ph}$.  When
$\kappa_{ph} = \kappa_0/3$, we see that the limit is achieved when
$\Delta$ is about $5k_B T$ ($E_g \approx 10 k_B T$), while for larger
$\kappa_{ph}$ it is achieved earlier at about $3k_B T$ for
$\kappa_{ph} = 10\kappa_0/3$ ($E_g \approx 6 k_B T$).  The values of
the so-called ``optimal band gap'' here are consistent with prior
theoretical studies, such as Refs.~\cite{sofo94-optgap}
and~\cite{hasdeo19}, in which the optimal band gaps are indicated to
exist within a range of $6$--$18~k_B T$.

By comparing the figure of merit from the two different values of $\kappa_{ph}$, we see that other than reducing the maximum $ZT$ value, a higher $\kappa_{ph}$ will also shift the band gap where the optimal $ZT$ is achieved. With higher $\kappa_{ph}$, the ratio $\sigma/\kappa$ becomes closer to $\sigma/\kappa_{ph}$, thus the chemical potential for maximum $ZT$ is pushed nearer to the band edge. The higher $\kappa_{ph}$ significantly decrease the $ZT$, but we note that $\kappa_{ph}$ falls off with temperature, while from eq.~\eqref{eq:kappafinal} the electron thermal conductivity should increase with temperature considering that the electron lifetime falls of by $T^{-3/2}$ at most. As such, at a higher temperature, the contribution of lattice thermal conductivity is then reduced and the thermoelectric performance of our material should be improved as the temperature is increased.

\subsection{Asymmetric two-band systems}
\label{subsec:asymm}

If previously in our model the conduction band and the valence band
have the same shape, now we would like to see the effects of different
conduction band and valence band shapes, i.e. when $m\neq n$.
Plotting the Seebeck coefficient and conductivities for $m=2$ and
$n=6$, we obtain Fig.~\ref{fig5}.

Since $S \propto S_c \sigma_c + S_v \sigma_v$ and $\sigma_c$ is
smaller than $\sigma_v$ while $S_c$, $S_v$ have opposite signs, we
obtain a higher total Seebeck coefficient in the case of asymmetric
bands [e.g. Fig.~\ref{fig5}(a) for $\Delta = 3 k_B T$ with $m = 2$ and
$n = 6$] than that in the case of symmetric bands
[e.g. Figs.~\ref{fig2}(a),~\ref{fig2}(d), and ~\ref{fig2}(g) for
$\Delta = 3 k_B T$ with $m = n = 2$, $m = n = 4$, and $m = n = 6$,
respectively].  The greatest increase is obtained when $m=2$ with a
peak in the Seebeck coefficient when chemical potential is well inside
the band gap.  On the other hand, since the values of conductivites at
the same chemical potential and at a particular value of band gap are
highly dependent on the ($m,n$) set, and for $m<n$ the conductivities
of the conduction band is lower than that of the valence band, our
calculated results for the asymmetric bands show that the total
conductivities [e.g. Figs.~\ref{fig5}(b-c) for $\Delta = 3 k_B T$ with
$m = 2$ and $n = 6$] are lowered compared to the previous case when
the bands are symmetric [e.g. Figs.~\ref{fig2}(b-c) for
$\Delta = 3 k_B T$ with $m = n = 6$].  The effects of increasing band
gap on the Seebeck coefficient and conductivities are shown in
Figs.~\ref{fig5}(d)--(f), consistent with the statements in the
previous subsection.

\begin{figure}[t]
  \centering\includegraphics[width=85mm,clip]{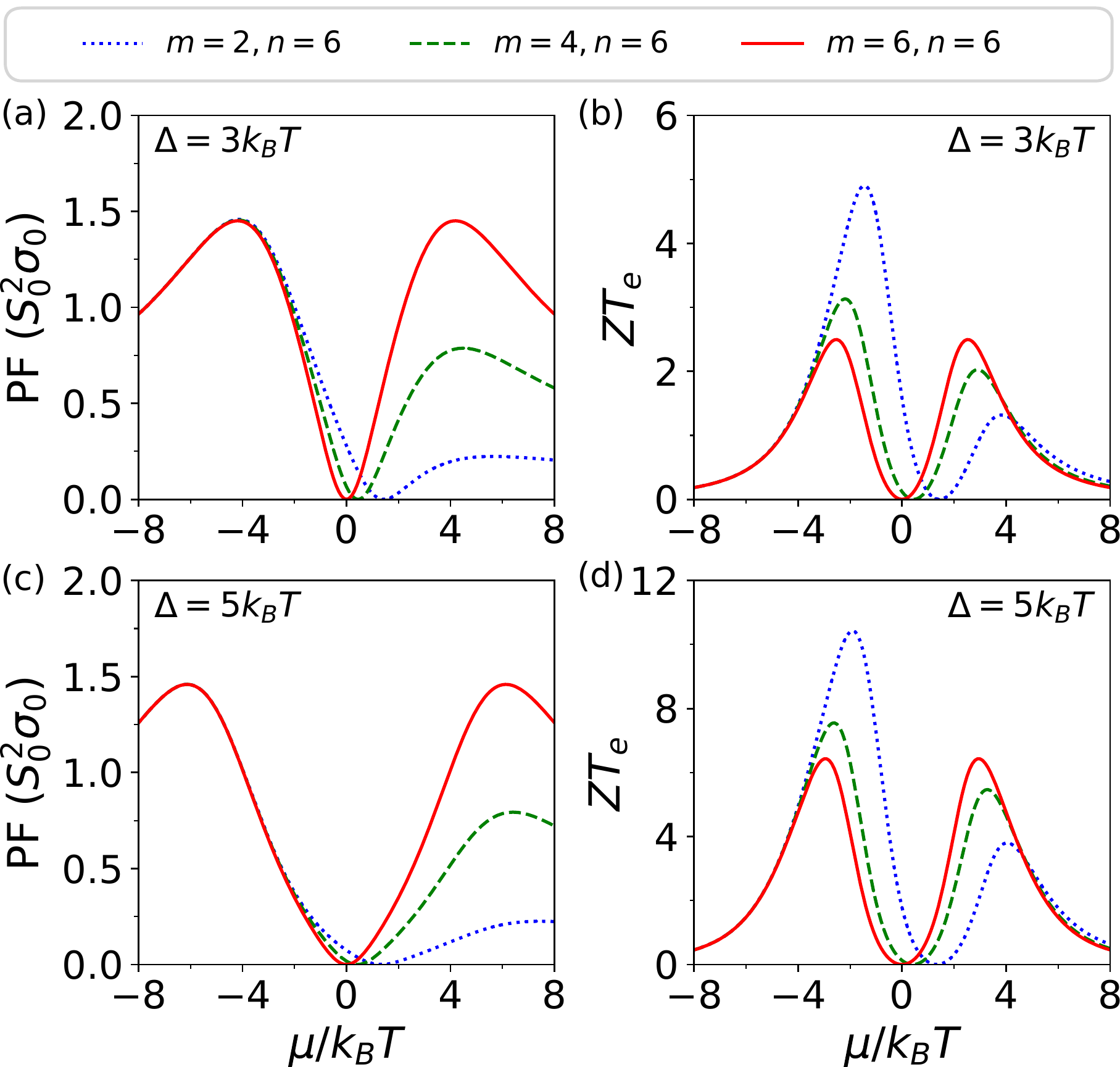}
  \caption{Power factors and electronic figures of merit considering
    asymmetric conduction bands with $m = 2, 4, 6$ and valence bands
    with $n = 6$.  Panels (a) and (b) are for to $\Delta = 3k_B T$
    ($E_g = 6 k_B T$), while (c) and (d) are for $\Delta = 5k_B T$
    ($E_g = 10 k_B T$).  Dotted, dashed, and solid lines indicate
    different combinations of $m$ and $n$ as shown in the top legend.}
  \label{fig6}
\end{figure}

While the increased Seebeck coefficient is beneficial, the reduced
conductivity can also reduce the power factor of the material.  We
plot the power factors and figures of merit for the cases of
$m < n$ and $m = n$ in Fig.~\ref{fig6} (note that as before, we
take $\kappa_{ph} = 0$).  However, we see that using $m < n$
(asymmetric) combination band does not increase the maximum possible
value of the power factor [Figs.~\ref{fig6}(a) and ~\ref{fig6}(c)]
because the peak of the Seebeck coefficient is obtained when the
chemical potential is still far from the band edge (inside the band
gap) while the electrical conductivity is still very low. 
On the other hand, the figure of merit can be significantly improved by the asymmetric band, provided that $\kappa_e \gg \kappa_{ph}$, as can be seen from Figs.~\ref{fig6}(b) and ~\ref{fig6}(d) which depict the ideal value for figure of merit when $\kappa_{ph} = 0$. As both electrical conductivity and electron thermal conductivity proportionally reduces for $m<n$, for $\kappa_e \gg \kappa_{ph}$ the ratio $\sigma/\kappa$ is relatively unchanged. This combination of unchanged $\sigma/\kappa$ and increasing $S$ leads to a potentially significant increase in the figure of merit.

\begin{figure}[t]
  \centering \includegraphics[width=85mm]{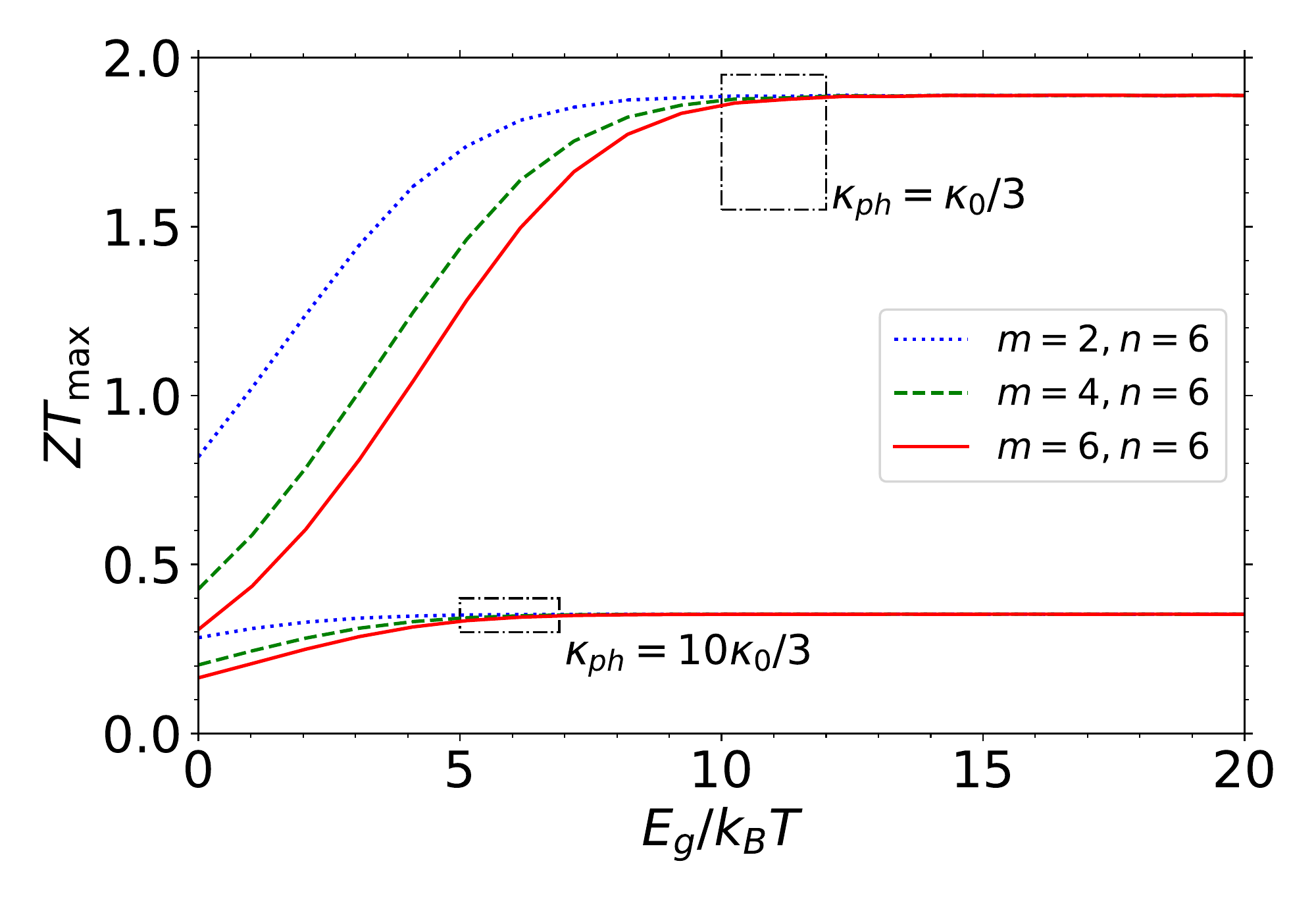}
  \caption{Maximum figure of merit $ZT_{\textrm{max}}$ (at the optimum
    chemical potential) as a function of band gap for several cases of
    asymmetric bands with two different parameters of phonon thermal
    conductivity: $\kappa_{ph} = \kappa_0/3$ and
    $\kappa_{ph} = 10 \kappa_0/3$.  Dash-dotted boxes serve as a guide
    for eyes to distinguish the calculated curves obtained from the
    different $\kappa_{ph}$ parameters.}
  \label{fig7}
\end{figure}

If we consider the effects of the band gap, we see in both
Figs.~\ref{fig5} and~\ref{fig6} that the qualitative behaviors of the
thermoelectric properties are similar to the symmetric case, with the
maximum power factor unaffected by either the asymmetric bands or the
increase in band gap.  On the other hand, the Seebeck coefficient and
figure of merit are significantly increased by both the asymmetric
band and the band gap.  We should note that for a real material the
increase in figure of merit will be much less dramatic, for in our
model, the decrease in electrical conductivity due to the smaller band
shape is always counteracted by a proportional decrease in thermal
conductivity.

In a real material with nonzero $\kappa_{ph}$, the thermal
conductivity will no longer be proportional to the electric
conductivity.  As a result, when the electron thermal conductivity is
low enough (for example due to high band gap) the improvement obtained
by the asymmetric bands in the figure of merit will begin to suffer.
To see the possible effects for non-zero phonon thermal conductivity,
as in the previous subsection we calculate the figure of merit with
$\kappa_{ph} = \kappa_0/3$ and $\kappa_{ph} = 10\kappa_0/3$.  We plot
the dependence of maximum figure of merit on band gap in
Fig.~\ref{fig7}.  Similar to our results from the previous subsection,
we see that as the band gap increases, $\kappa_e$ becomes smaller than
$\kappa_{ph}$ and $ZT \approx S^2 \sigma/\kappa_{ph} \propto PF$,
where $\kappa_{ph}$ is assumed to be independent of band gap.  When
this point is achieved, the peak of $ZT$ is located at the region near
a band edge where the effect of the opposing band is negligible.
Therefore, increasing the band gap will no longer increase the figure
of merit, and at this point, there will be no appreciable difference
between the symmetric and asymmetric bands because when the band gap
is large enough the two-band model effectively becomes a one band
model where the effect of the opposing band is negligible, whatever
its shape is.  That being said, for narrow-gap materials the presence
of asymmetric bands can achieve higher $ZT$ values than the symmetric
bands.  These maximum figures of merit are obtained at a lower doping
level, thus this kind of material might have an advantage in ease of
manufacturing.

\begin{figure*}[t]
  \centering \includegraphics[width=16cm,clip]{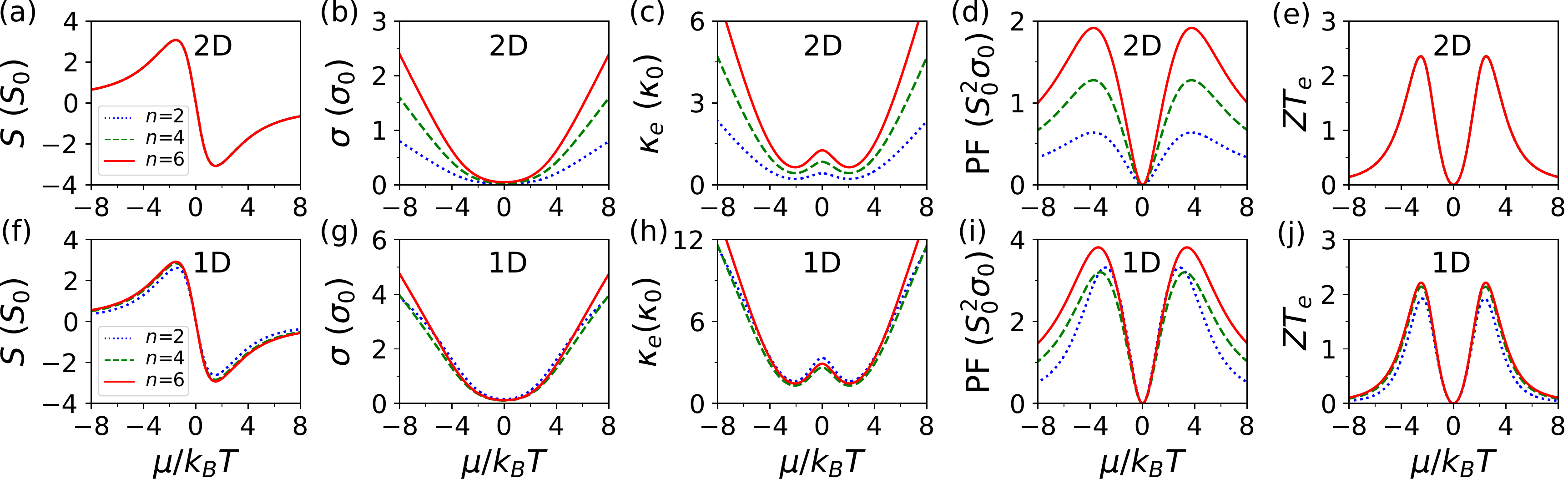}
  \caption{Thermoelectric properties of (a--e) 2D and (f--j) 1D
    symmetric-band $(m = n)$ systems with $\Delta = 3k_B T$.
    Calculated results for different powers of dispersion, $n = 2, 4$,
    and $6$, are represented by dotted, dashed, and solid lines,
    respectively.}
    \label{fig8}
\end{figure*}

\subsection{Effects of dimensionality}
\label{subsec:dim}

\begin{figure*}[t]
  \centering\includegraphics[width=16cm]{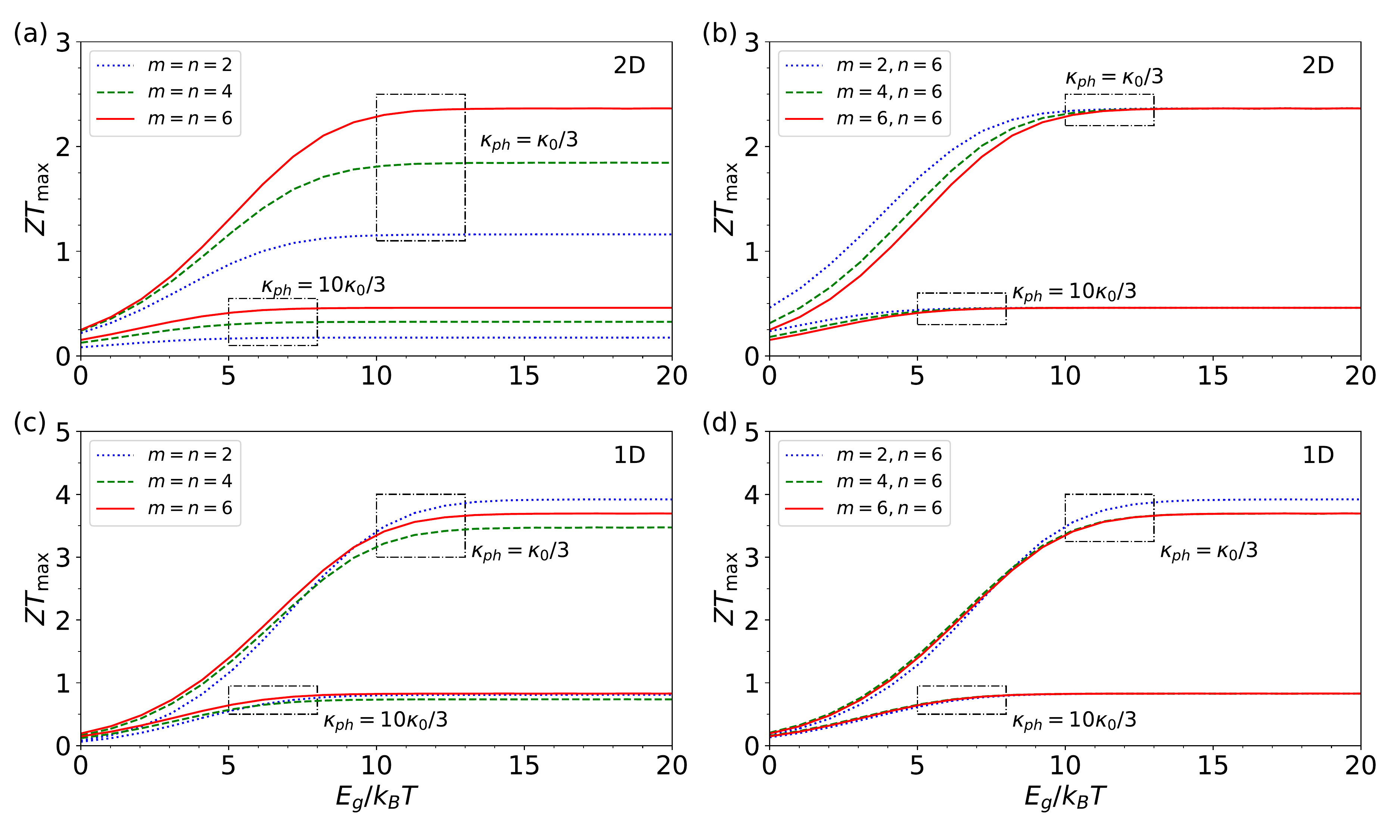}
  \caption{Maximum figure of merit $ZT_{\textrm{max}}$ (at the optimum
    chemical potential) as a function of band gap for (a) 2D symmetric
    ($m = n$) bands, (b) 2D asymmetric ($m \neq n$) bands, (c) 1D
    symmetric bands, and (d) 1D asymmetric bands.  The values of $m$
    and $n$ are given in the legend of each panel.  Two different
    parameters of phonon thermal conductivity are considered, i.e.
    $\kappa_{ph} = \kappa_0/3$ and $\kappa_{ph} = 10 \kappa_0/3$.
    Dash-dotted boxes serve as a guide for eyes to distinguish the
    calculated curves obtained from the different $\kappa_{ph}$
    parameters. }
    \label{fig9}
\end{figure*}

We have not discussed yet the effects of the material's dimensionality on
the figure of merit.  The most important effect of the dimension on the
thermoelectric properties of a semiconductor is the different DOS for
each dimension as previously shown in Fig.~\ref{figTEDOS}, leading to
a difference in the prefactor for the electrical and thermal
conductivities and an extra factor in the $\mathcal{G}_i$
integrals~[Eqs.~\eqref{eq:Gc} and~\eqref{eq:Gc}].  Here we note that
although the density of states diverges near the band edge when $n>2$
(and when $n=2$ too for the 1D case), the resulting thermoelectric
integral is always finite because the group velocity goes to zero near
the band edge.  Using the formulas derived in Section~\ref{sec:th},
with $d = 2$ and $d = 1$, we plot the thermoelectric properties of 2D
and 1D materials for $\Delta = 3k_B T$ in Figs.~\ref{fig8}(a--e) and
Figs.~\ref{fig8}(f--j), respectively.

From Eqs.~\eqref{eq:Sigmafinal} and \eqref{eq:kappafinal}, due to the
presence of the factors involving the number of dimensions $d$ 
in the longitudinal velocity $v$, reduction of $d$ will
increase the electrical and thermal conductivities.  As the power
factor is proportional to the electrical conductivity while the figure
of merit is proportional to $\sigma/\kappa$ only, we can conclude that
the reduction of $d$ increases the power factor but only barely
affects the figure of merit.  We see some slight differences in the
behavior of the Seebeck coefficient for different dimensions.  The
slight difference results from the presence of the
$(x+\eta-\Delta)^{(d-2)/n}$ factor in the $\mathcal{G}_i$ integrals.
For the 2D case, this factor completely vanishes, which means the
Seebeck coefficient is completely independent of dispersion power $n$
while for the 1D case, this extra factor becomes
$(x+\eta-\Delta)^{-1/n}$, which increases the Seebeck coefficient
while $n$ is higher, unlike the 3D case.  One interesting phenomenon
that can be observed from these plots is the crossover of the
electrical and thermal conductivities in the 1D case between $n=2$ and
$n=4,n=6$.  This crossover occurs because near the band edge the group
velocity is larger for smaller values of $n$, while far from the band
edge the group velocity becomes smaller for large values of $n$
instead.  This happens for all dimensions; however, in the 2D and 3D
cases, the DOS for the $n=2$ band is lower than the $n=4$ and $n=6$
bands near the band edge, as can be seen from Figs.~\ref{figTEDOS}(b)
and~\ref{figTEDOS}(c).  Meanwhile, for the 1D case
Fig.~\ref{figTEDOS}(a), similar to the DOS of $n=4$ and $n=6$ bands,
the DOS of the $n=2$ band diverges.

\begin{figure*}[t]
  \centering \includegraphics[width=16cm]{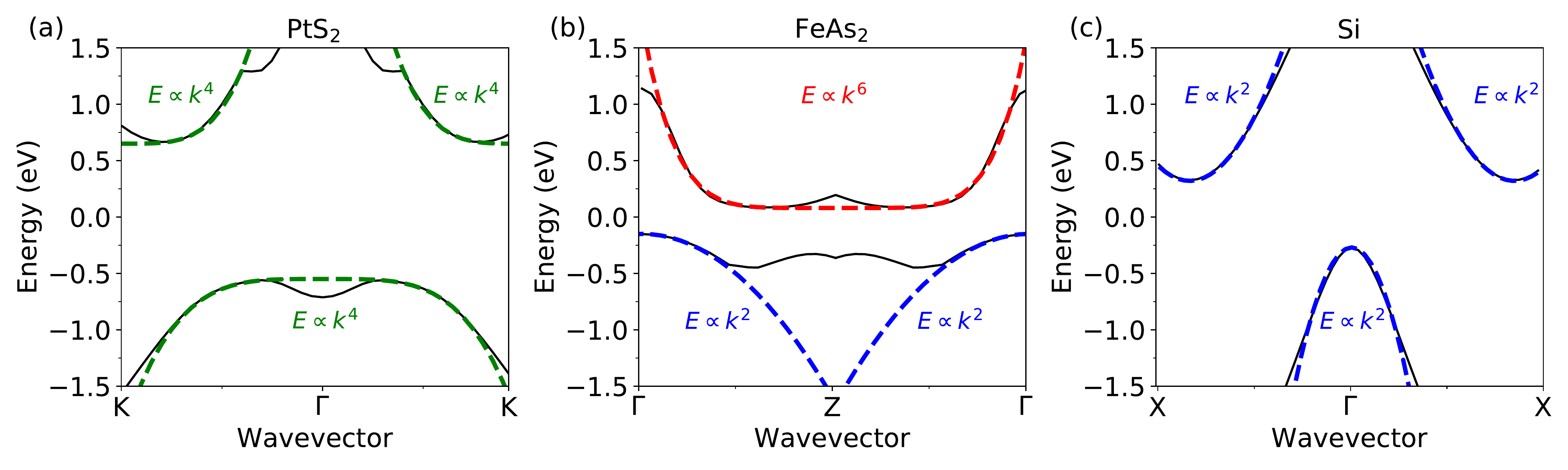}
  \caption{Fitting the conduction and valence bands of
    (a)~$\mathrm{PtS}_2$, (b)~$\mathrm{FeAs}_2$, and (c)~Si.
    Depending on shapes of the bands, we may fit them to pudding-mold
    ($E \propto k^4$ and $E \propto k^6$) or parabolic
    ($E \propto k^2$) energy dispersion.}
  \label{figTEfit}
\end{figure*}

\begin{figure}[t]
  \centering \includegraphics[width=85mm]{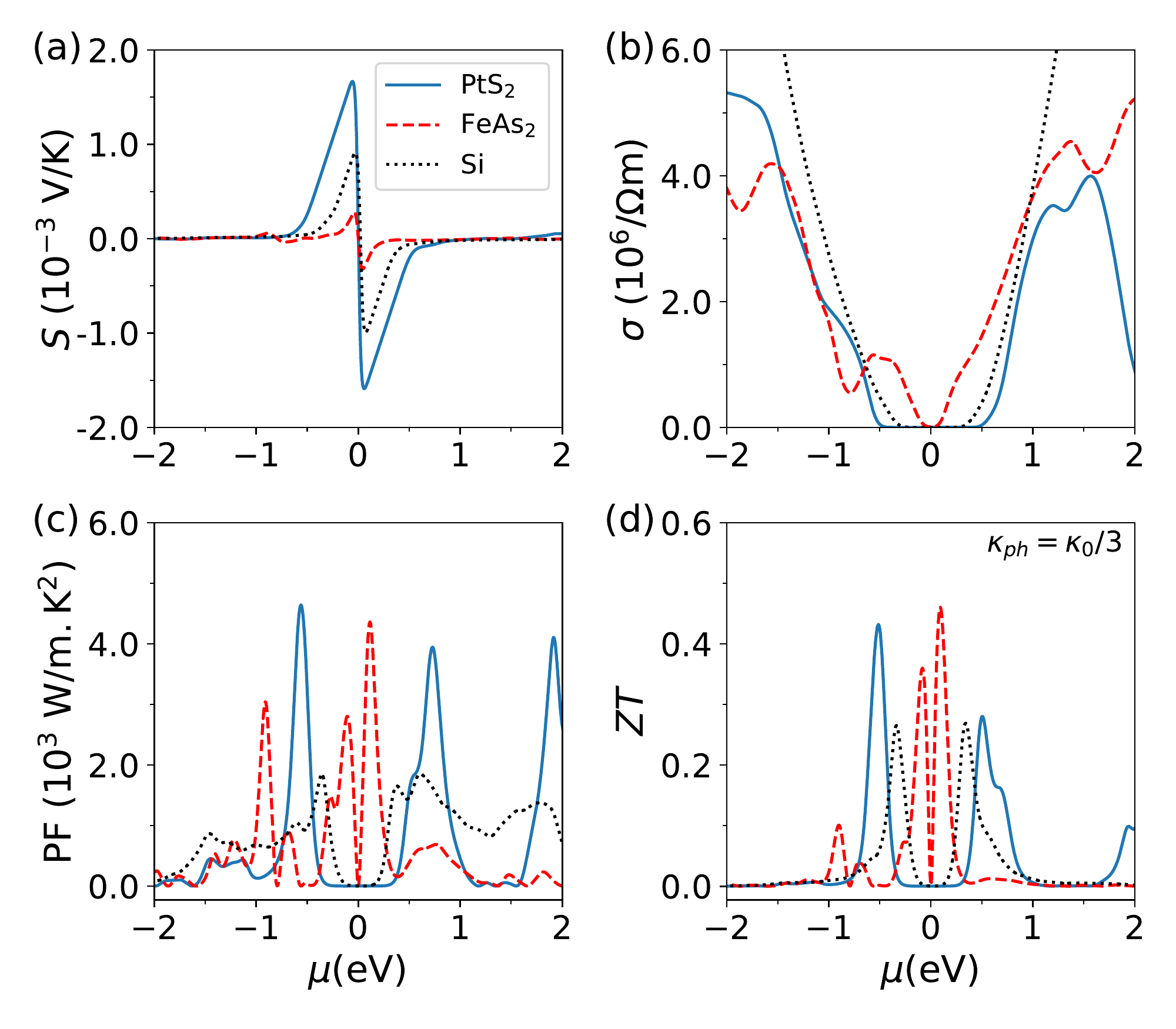}
  \caption{Calculated thermoelectric properties of $\mathrm{PtS}_2$,
    $\mathrm{FeAs}_2$, and Si represented by solid, dashed, and dotted
    lines, respectively.  (a)~Seebeck coefficient, (b)~electrical
    conductivity, (c)~power factor, and (d)~dimensionless figure of
    merit.}
  \label{FigTEPFreal}
\end{figure}

We show the plots of $ZT_{\mathrm{max}}$ versus band gap for the 2D
and 1D systems in Fig.~\ref{fig9}, with $\kappa_{ph} = \kappa_0/3$ and
$10 \kappa_0/3$.  We also consider the cases of $m = n$
[Figs.~\ref{fig9}(a) and~\ref{fig9}(c)] and
$m \neq n$~[Figs.~\ref{fig9}(b) and~\ref{fig9}(d)].  For the 2D
systems [Figs.~\ref{fig9}(a) and~\ref{fig9}(b)], the behavior of the
figure of merit is essentially similar to the 3D case, in which there
is an increase in $ZT_{\mathrm{max}}$ as the band gap is increased up
to a certain limit.  As depicted in Fig.~\ref{fig9}(a), the
pudding-mold bands give a better $ZT$ than the parabolic band when
$m = n$.  On the other hand, in Fig.~\ref{fig9}(b), we see that the
asymmetric ($m \neq n$) bands give a higher $ZT$ than the symmetric
band at lower band gaps.  The overall $ZT$ for the 2D case is slightly
larger than that for the 3D case due to the presence of a larger
dimension-dependent factor ($C_d/d$) for $\sigma$ and $\kappa_e$.

For the 1D systems [Figs.~\ref{fig9}(c) and ~\ref{fig9}(d)], with
nonzero $\kappa_{ph}$ we surprisingly find that $ZT_{\mathrm{max}}$
values at certain band gaps for the cases containing parabolic bands
are actually higher than those with pudding-mold band.  This behavior
is due to the higher electrical and electron thermal conductivities
for regions near the band gap in the parabolic band.  This higher
conductivities lead to a higher $\sigma/\kappa$. The presence of the
dimension-dependent factor $C_d/d$ also causes the figure of merit for
the 1D cases [Figs.~\ref{fig9}(c) and ~\ref{fig9}(d)] to be higher
than the figure of merit for the 2D cases [Figs.~\ref{fig9}(a) and
~\ref{fig9}(b)].

\subsection{Insight into real materials}
\label{subsec:real}

To get the insight into real materials from our model so far, we
calculate the Seebeck coefficient, electrical conductivity, power
factor, and figure of merit of some semiconductors which are expected
to contain pudding-mold bands as well as parabolic bands.  At least,
we want to see more realistic effects when we have different band gaps
and band shapes (dispersion powers $n$), so that we can suggest a
better direction for finding potential thermoelectric materials from
the simple model described before.

Here we consider platinum sulfide ($\mathrm{PtS}_2$) as a
representative of large-gap semiconductors, iron arsenide
($\mathrm{FeAs}_2$) as a representative of narrow-gap semiconductors,
and silicon (Si) as a commonly-used semiconductor in daily life.
We take $\mathrm{FeAs_2}$ and $\mathrm{PtS_2}$ to represent pudding-mold materials since $\mathrm{FeAs_2}$ has previously been noted to exhibit a quasi one dimensional pudding-mold band~\cite{kuroki13}. Meanwhile, another platinum-chalcogen compound, $\mathrm{PtSe_2}$, has been noted as a good thermoelectric material with a pudding-mold band~\cite{usui2014}, thus we expect $\mathrm{PtS_2}$ to have similar properties.
We calculate the band structure of these materials using the Quantum
ESPRESSO package~\cite{QE-2017} in their 3D forms.  Then, with the
information about the band structure, we calculate their Seebeck
coefficients, electrical conductivities, and electron thermal
conductivities using the BoltzTraP
package~\cite{madsen2006boltztrap,BoltzTraP2}.  We take the crystal
structure data of these materials and the suggested parameters for
Quantum ESPRESSO calculations from the AFLOWLIB
repository~\cite{aflow2012,aflowlib2012}.  For simplicity, we do not
calculate the phonon thermal conductivity, but use an estimated value
corresponding to $\kappa_{ph} = \kappa_0 / 3$, which is quite
realistic for these materials.  Note that there is importance of using a two-band model to explain TE properties of narrow gap materials such as $\mathrm{FeAs_2}$. For many materials, one may obtain the optimal TE properties when the Fermi energy is near the band edge, i.e., when the doping is relatively small \cite{Li2017, Hinterleitner2019, Zhang2017}. As a result, when calculating the TE properties of narrow-gap materials, we cannot neglect the contribution of bipolar effects. Indeed, as shown through our calculation, the two-band model can capture the effects of asymmetric valence-conduction band structures on the reduction of bipolar effects. This result suggests that for narrow-gap materials the optimal TE performance can be achieved without having a high doping level.

The band structures of $\mathrm{PtS}_2$, $\mathrm{FeAs}_2$, and Si along certain high-symmetry points are shown in Figs.~\ref{figTEfit}(a),~\ref{figTEfit}(b), and~\ref{figTEfit}(c).  See also \ref{append:dft} for more details of the band structures.  Note that the indirect gap features in the band structures do not affect the thermoelectric properties considered in our model because the contributions from each band, subband, or valley to the thermoelectric transports can be independently separated and summed up~\cite{pei2011convergence}.  For $\mathrm{PtS_2}$ at the $K$--$\Gamma$ path, as shown in Fig.~\ref{figTEfit}(a), we see that the conduction and valence band can be fitted with a quartic band $m=n=4$.  Next, in Fig.~\ref{figTEfit}(b), we see that $\mathrm{FeAs_2}$ has a wide pudding-mold conduction band along the $\Gamma$--$Z$ points that which we fit with a sextic band $m=6$ while we fit the rounded edge of the valence band nearby with a parabolic band. We note that $\mathrm{FeAs_2}$ also has a very wide, almost flat band around the R-T-Z and Y-T direction, but following the argument in Ref.~\cite{kuroki13}, we can treat this anisotropy as causing the pudding-mold band to be effectively one-dimensional in the dispersive direction.  In the case of Si, we simply fit the conduction and valence bands by parabolic $(n = 2)$ forms.

We show the Seebeck coefficient, electrical conductivity, power
factor, and figure of merit in Fig.~\ref{FigTEPFreal}. Comparing the
band gap of the three different materials, we see that since the band
gap of $\mathrm{FeAs}_2$ is much lower than $\mathrm{PtS_2}$ and
$\mathrm{Si}$, the Seebeck coefficient of $\mathrm{FeAs}_2$ is
significantly lower [Fig.~\ref{FigTEPFreal}(a)], but the electrical
conductivity around $\mu = 0$ is much higher than the other two
materials which are almost zero [Fig.~\ref{FigTEPFreal}(b)].  We also
observe that $\mathrm{PtS}_2$ has a higher Seebeck coefficient
compared to $\mathrm{Si}$ due to its slightly larger band gap.  
From the results of electrical conductivity [Fig.~\ref{FigTEPFreal}(b)], we
observe that $\mathrm{PtS}_2$ and $\mathrm{Si}$ have comparable
conductivities near the band edges although the band gap of
$\mathrm{PtS}_2$ is larger than the band gap of $\mathrm{Si}$.  We can
attribute this behavior to the presence of pudding-mold band for
$\mathrm{PtS}_2$ in the valence band with $n = 4$ since which there
exists a sharp increase in the conductivity of $\mathrm{PtS}_2$ when
$\mu = 0.6$--$0.7~\mathrm{eV}$.  

Next, for the power factor, we see that both $\mathrm{PtS}_2$ and
$\mathrm{FeAs}_2$ achieve higher peaks [Fig.~\ref{FigTEPFreal}(c)] than
$\mathrm{Si}$, with the best peak of $\mathrm{FeAs}_2$ slightly higher
than $\mathrm{PtS}_2$.  Since from Figs.~\ref{fig3}(a--c) we know that
the maximum power factor should be relatively unaffected by band gap
and we see that $\mathrm{FeAs}_2$ has a comparable power factor to
$\mathrm{PtS}_2$ despite its much lower band gap, we argue that the
higher power factors (compared to Si) obtained in $\mathrm{FeAs}_2$
and $\mathrm{PtS}_2$ are caused by the improvement in the electrical
conductivity due to the presence of pudding-mold bands compared to a
material without pudding-mold bands. 

Consistent with results of power factor, as shown in
Fig.~\ref{FigTEPFreal}(d), the $ZT$ values for $\mathrm{PtS}_2$ and
$\mathrm{FeAs}_2$ are better than those for Si in a wide range of
$\mu$. It is interesting that the $ZT$ and power factor for $\mathrm{FeAs_2}$ is not significantly smaller than that of $\mathrm{PtS_2}$, even though the low band gap of $\mathrm{FeAs_2}$ implies the presence of a significant bipolar effect. Considering the band sturcture of $\mathrm{FeAs}_2$ from Fig.~\ref{figTEfit}(b), it is reasonable to conclude that the asymmetrical nature of the band structure, along with the presence of $m=6$ conduction band, are enough to counteract the reduction of $PF$ and $ZT$ due to bipolar effects. The lower $PF$ and $ZT$ for lower doping level is also consistent with this hypothesis. However, we note that the $ZT$ values for $\mathrm{PtS}_2$ and
$\mathrm{FeAs}_2$, even when using $\kappa_{ph} = \kappa_0 / 3$ (not
$\kappa_{ph} = 10 \kappa_0 / 3$), are still less than 1, indicating
that further search or screening of real materials with the
pudding-mold band structures are desirable for obtaining better $ZT$
values because as explained in
Secs.~\ref{subsec:symm}--\ref{subsec:dim}, consideration of the
pudding-mold band structures hold promise for $ZT > 1$.  We expect
that near future works on the high-throughput calculations for the
materials search, in all possible dimensions, can be triggered by this
present work.

\section{Conclusions}
\label{sec:con}

We find that pudding-mold type (quartic and sextic) bands provide both
higher power factors and figures of merit compared to parabolic bands
in 2D and 3D materials due to the increased conductivities, while in
1D material the parabolic bands may have larger figures of merit than
the pudding-mold bands at a certain range of optimal band gaps.  Using
two bands with different shapes in the conduction and valence bands
can also achieve an improvement in figure of merit at a lower band gap
and chemical potential.  Comparing the results for 3D, 2D, and 1D
materials, we see that decreasing the dimension does generally
increase the figure of merit, but it should be noted again that for a
1D material there is no significant improvement in using the
pudding-mold bands over the parabolic bands due to the similarity in
the DOS of each band.

\section*{Data availablility}
The raw/processed data required to reproduce these findings are available to download from \url{http://github.com/jyestama/PuddingTE}.

\section*{CRediT authorship contribution statement}
\textbf{J.A. Adhidewata:} Software, Formal analysis, Investigation, Data Curation, Writing - original draft, Visualization. \textbf{A.R.T. Nugraha:} Conceptualization, Methodology, Validation, Writing - review \& editing, Visualization, Supervision. \textbf{E.H. Hasdeo:} Conceptualization, Methodology, Validation, Writing - review \& editing.\textbf{ P.E. Estell\'e:} Writing - review \& editing.\textbf{ B.E. Gunara:} Writing - review \& editing, Supervision, Project administration, Funding acquisition.

\section*{Declaration of Interests}
The authors declare that they have no known competing financial
interests or personal relationships that could have appeared to influence
the work reported in this paper.

\section*{Acknowledgements}
  We acknowledge Mahameru LIPI for their high-performance computing
  facilities.  A.R.T.N. also acknowledges SSHN fellowship from the
  French Embassy in Indonesia for a visit to U. Rennes.  J.M.A. and
  B.E.G. are supported by P3MI ITB and P2MI ITB.

\appendix

\section{Some notes on the bandwidth parameter}
\label{append:bandwidth}
\begin{figure*}[t]
  \centering \includegraphics[width=16cm,clip]{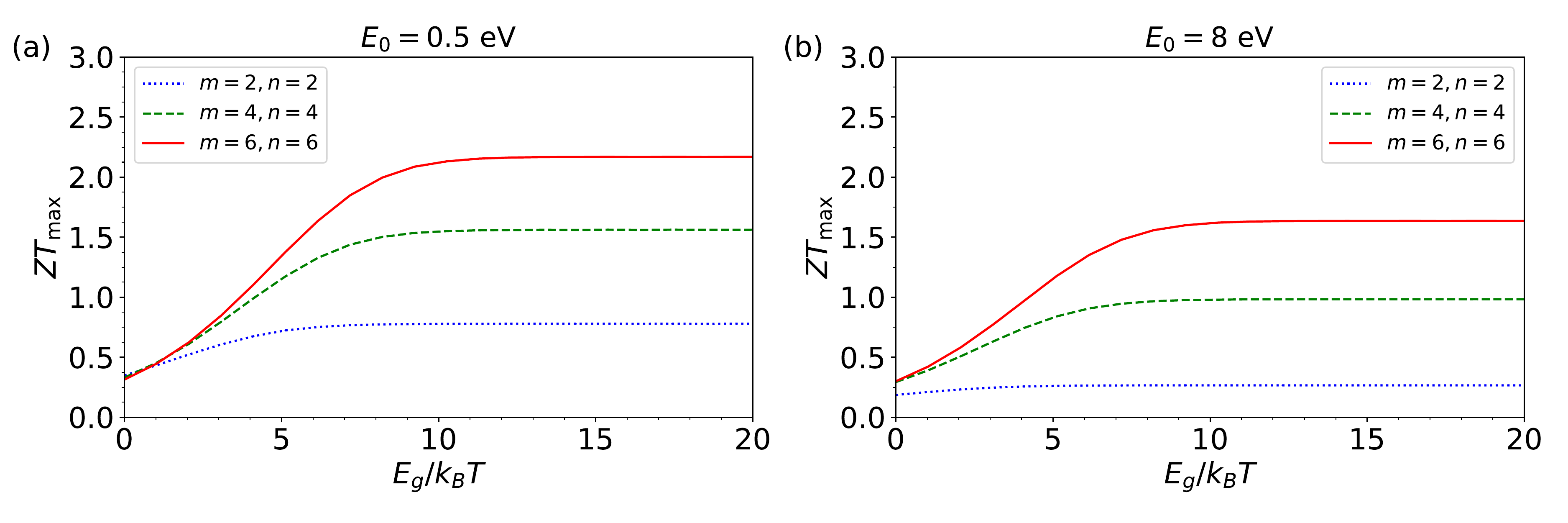}
  \caption{The figure of merit for a 3D material with (a) $E_0=0.5$ eV and (b)$E_0=8$ eV as a function of band gap. All other parameters are the same as in the main text}
  \label{FigC1}.
\end{figure*}

For all of our calculations, the coefficient $A_n$ in the band dispersion relation is given by
\begin{equation}
  A_n = \left(\frac{a}{\sqrt{d}\pi}\right)^n E_0,
\end{equation}
where we take $E_0$ to be the same in all cases, $E_0 = 2 \mathrm{eV}$. By choosing this form of $A_n$ we ensure that for all values of $n$ the bands will converge in the edge of the (cubic) first Brillouin zone, at $E = E_0$. Here we examine the dependence of the TE properties on the value of $E_0$.  By increasing $E_0$, the dispersion curve will become steeper, which will increase the electron velocity at the cost of reducing the density of states. From Eqs.~(\ref{eq:Sigmafinal})-(\ref{eq:kappafinal}) we see that the band width affects the electrical and electron thermal conductivity as a multiplicative factor which also depends on the dimension and $n$. For a 3D material, the reduction in density of states exceeds the increase in band velocity and we found that $\sigma, \kappa_e \sim (k_B T/E_0)^{1/n}$, while the Seebeck coefficient is independent of $E_0$. This implies that a larger $E_0$ will lead to a smaller conductivity, which means an increase in $E_0$ will likely decrease the power  factor and figure of merit, and vice versa. Due to the $1/n$ exponent, this change in conductivity will be more pronounced in parabolic bands compared to pudding-mold bands. We plot the resulting figure of merit in 3D materials for two different values of $E_0$ in Fig.~\ref{FigC1}. We see that while this change in $E_0$ slightly increases or decreases the figure of merit as expected, the general trend and the relation between band shape and figure of merit is similar to that of the main result. 

\begin{figure*}[t]
  \centering \includegraphics[width=16cm,clip]{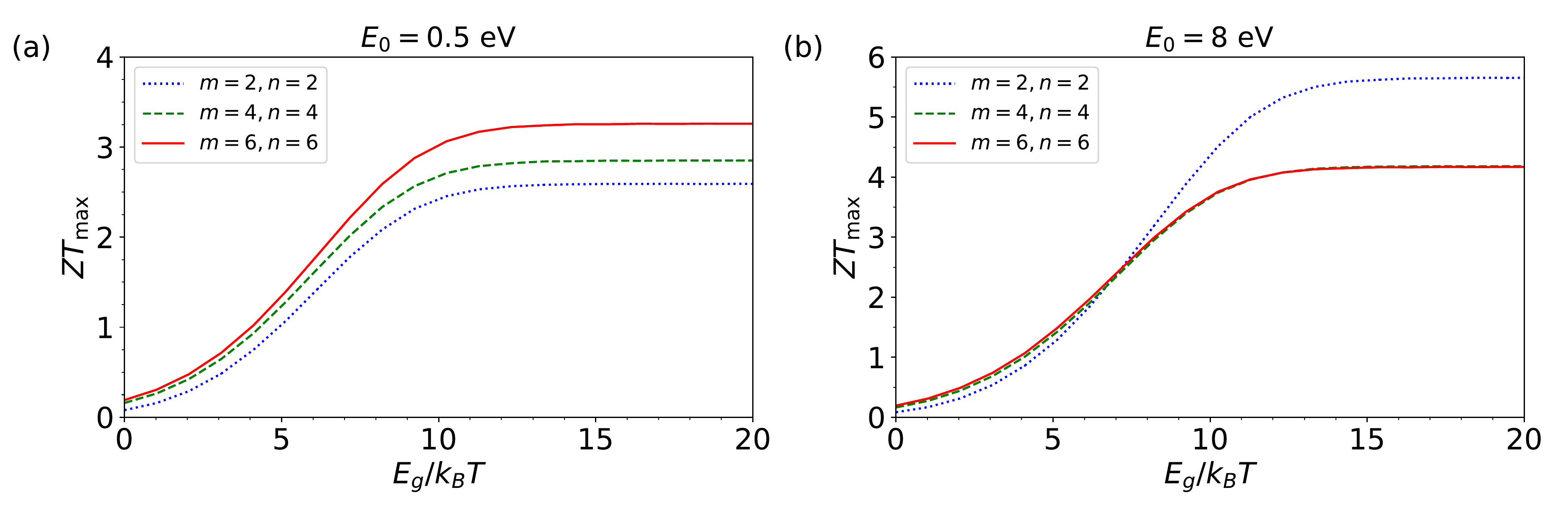}
  \caption{The figure of merit for a 1D material (a) $E_0=0.5$ eV and (b)$E_0=8$ eV as a function of band gap. All other parameters are the same as in the main text}
  \label{FigC2}.
\end{figure*}

From Eqs.~(\ref{eq:Sigmafinal})-(\ref{eq:kappafinal}) it is apparent that the conductivities for 2D materials are not affected by the bandwidth. Meanwhile, for 1D materials, the increase in velocity obtained by increasing $E_0$ is larger then the decrease in density of states; we see that $\sigma, \kappa_e \sim (E_0/k_B T)^{1/n}$ which means an increase in $E_0$ will increase the power factor and figure of merit instead. We plot the figure of merit in 1D materials for two different values of $E_0$ (similar to the previous plot) in Fig.~\ref{FigC2}. We see that increasing (decreasing) $E_0$ improves (decreases) the figure of merit, with a larger change observed for smaller $n$. Notably, for a low enough $E_0$ the figure of merit of the pudding-mold bands beat the figure of merit of the parabolic band, unlike the result in the main text, while for bigger $E_0$ the improvement in performance for the parabolic band compared to the pudding mold bands becomes more significant. 

\section{Calculation with DOS-dependent relaxation time approximation (DRTA)}
\label{append:drta}
\begin{figure*}[t]
  \centering \includegraphics[width=16cm,clip]{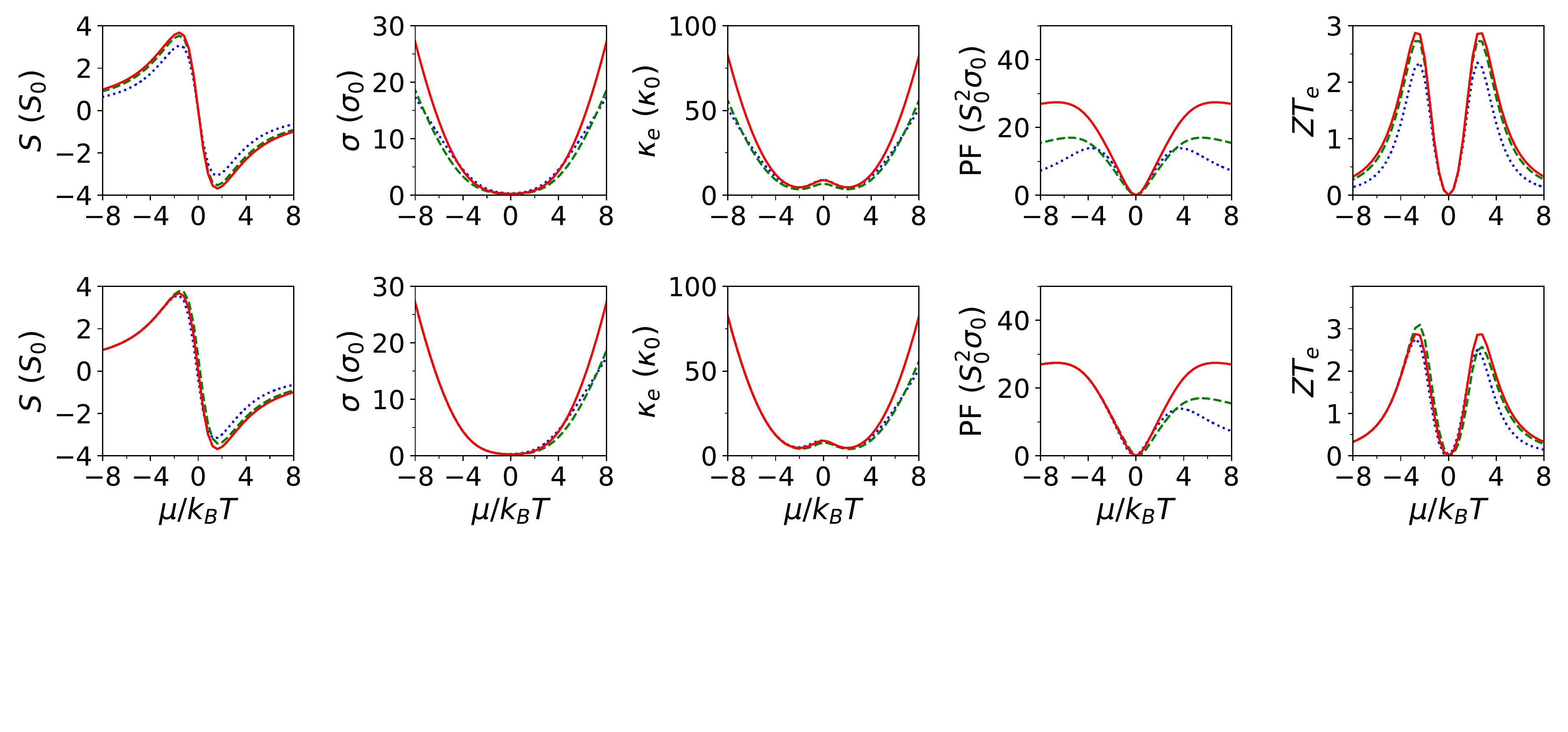}
  \caption{Thermoelectric properties of  (a--e) symmetric-band $(m = n)$ and (f--j) asymmetric band systems $m\neq n$ with the $\tau \sim \mathrm{DOS}^{-1}$ approximation. We take $\Delta = 3k_B T$.
    Calculated results for different powers of dispersion, $n = 2, 4$,
    and $6$, are represented by dotted, dashed, and solid lines,
    respectively.}
    \label{figA1}
\end{figure*}

Using similar methods as described in Sec.~\ref{sec:th}, we calculate the TE properties of our pudding-mold band model by taking the relaxation time to be
\begin{equation}
    \tau = C/g(\epsilon)
\end{equation}
Substituting this value of $\tau$ to Eqs. \eqref{eq:Lic} and \eqref{eq:Liv}, we obtain
\begin{equation}
\mathcal{L}_{c,i} = \frac{Ca^2 (k_BT)^{i+2}}{d^2 \pi^2 \hbar^2} \frac{1}{(k_B T/E_0)^{2/n}} n^2 \mathcal{G}_{c,i}^{1/DOS}(\eta,\tilde{\Delta}),
\end{equation}
where
\begin{equation}
    \mathcal{G}_{c,i}^{1/DOS}(\eta,\tilde{\Delta}) = \int_{\Delta'-\eta}^{\infty} (x+\eta-\Delta)^{2-2/n} \frac{e^x}{(e^x+1)^2} (x)^i dx.
\end{equation}
A similar relation for the valence band in analogy with \eqref{eq:Gv} can also be obtained. We then plot the results for the Seebeck coefficient, electrical conductivity, electron thermal conductivity, power factor, and ideal figure of merit in Figs.~\ref{figA1}. Here the electrical conductivity and electron thermal conductivity are expressed in the constants:
\begin{align}
    \sigma_0^{1/DOS} &= \frac{Ce^2 a^2 k_B T}{\hbar^2}, \\
    \kappa_0^{1/DOS} &= \frac{Ca^2 k_B^4 T^3}{\hbar^2}.
\end{align}

\begin{figure*}[t]
  \centering\includegraphics[width=16cm,clip]{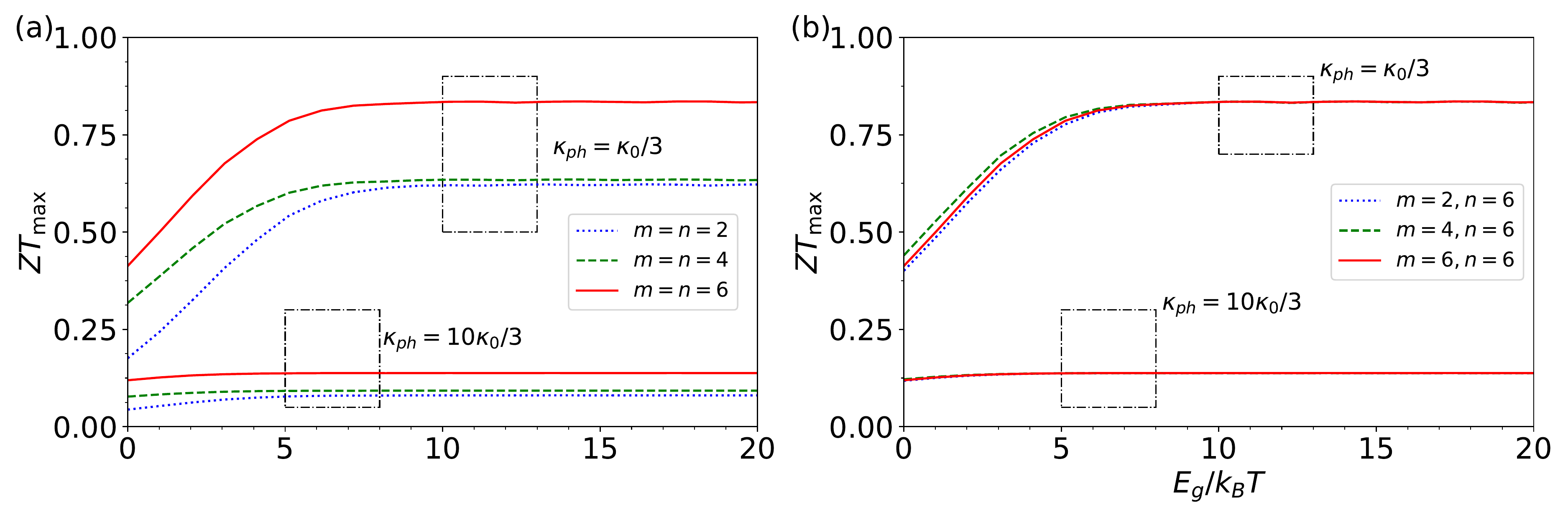}
  \caption{Maximum figure of merit $ZT_{\textrm{max}}$ (at the optimum
    chemical potential) as a function of band gap for (a) symmetric
    ($m = n$) bands, (b) asymmetric ($m \neq n$) bands with $\tau \sim (\mathrm{DOS})^{-1}$.  The values of $m$
    and $n$ are given in the legend of each panel.  Two different
    parameters of phonon thermal conductivity are considered, i.e.,
    $\kappa_{ph} = \kappa_0/3$ and $\kappa_{ph} = 10 \kappa_0/3$.
    Dash-dotted boxes serve as a guide for eyes to distinguish the
    calculated curves obtained from the different $\kappa_{ph}$
    parameters. }
    \label{figA2}
\end{figure*}

In Figs.~\ref{figA2}(a)--(b), we also plot the optimum figure of merit $ZT_{max}$ for each value of band gap, with two different possible phonon thermal conductivity. As in the main text, we assume the phonon thermal conductivity to be constants in the realistic range of $1$--$10 \mathrm{W/m K}$. We note that the dimension dependence of $\mathcal{L}_i$s is only in the form of a uniform $1/d^2$ scaling factor, as such we will only plot the results for the 3D material.   

While overall the results in Fig.~\ref{figA1} seem similar to the results from the main text, there is one important difference that can be clearly seen from Fig.~\ref{figA2}. The dependence of the band gap on the figure of merit looks the same, but unlike our previous result, we see a similar value of $ZT$ for $n=2$ and $n=4$. This relative increase in the $n=2$ figure of merit can be explained if we take a closer look at the electrical conductivity. Because for the DRTA approximation the DOS factor in the thermoelectric integrals \eqref{eq:Lic}, \eqref{eq:Liv} vanishes, the conductivity only depends on the velocity. Near the band edge, the pudding mold bands have a flat shape, thus for low chemical potential, the conductivity of the parabolic band is higher than that of the $n=4$ band. On the other hand, in the CRTA approximation, this nearly zero velocity is partially counteracted by the singular DOS at the band edge. This lower conductivity for the $m=4$ band compared to the $m=2$ also means that for the asymmetrical band configuration, the $m=4,n=6$ case provides a slightly better result than the $m=2, n=6$ shape.

\section{Band structure calculations}
\label{append:dft}
\begin{figure}[t]
  \centering \includegraphics[width=85mm]{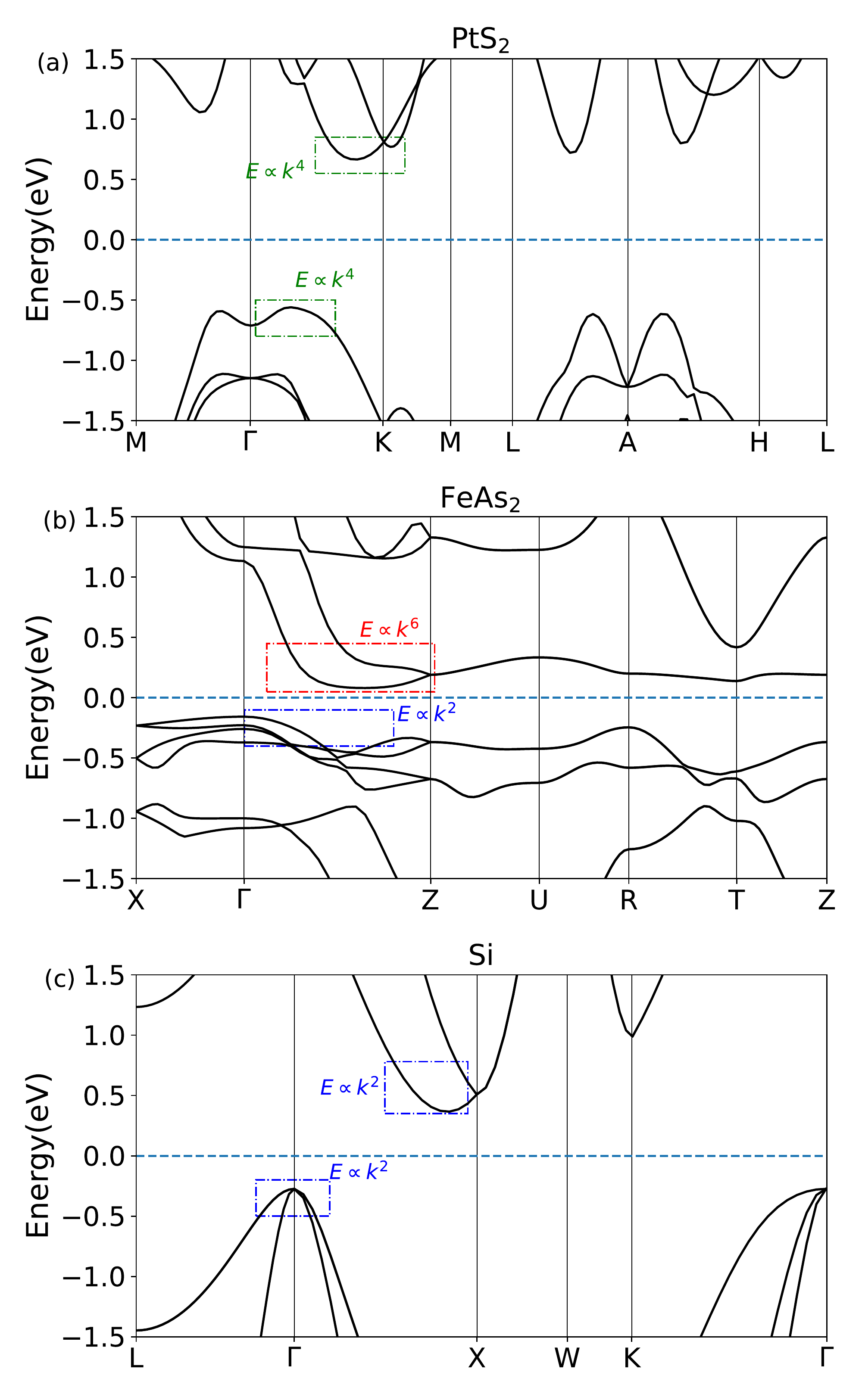}
  \caption{The band structure for (a) $\mathrm{PtS_2}$, (b) $\mathrm{FeAs_2}$, and (c) $\mathrm{Si}$ obtained from first principles calculation. Dashed boxes show the position of pudding-mold bands that we plot in Fig.~\ref{figTEfit}}
  \label{Figbandsreal}.
\end{figure}

In Figs.~\ref{Figbandsreal}(a)--(c), we show the complete band structures of $\mathrm{FeAs_2}, \mathrm{PtS_2},$ and $\mathrm{Si}$ that we obtain from first principles Quantum ESPRESSO calculations. In our calculations we use projected augmented wave (PAW) pseudopotentials taken from PSLibrary~\cite{pslibrary, dalcorso2014}. To ensure a reasonable computation time, we set the kinetic energy cutoff to be 60 Ry for $\mathrm{FeAs_2}$ and $\mathrm{Si}$, and 100 Ry for $\mathrm{PtS_2}$. The convergence threshold is set to $10^{-8}$ Ry.



\begin{thebibliography}{10}
\expandafter\ifx\csname url\endcsname\relax
  \def\url#1{\texttt{#1}}\fi
\expandafter\ifx\csname urlprefix\endcsname\relax\def\urlprefix{URL }\fi
\expandafter\ifx\csname href\endcsname\relax
  \def\href#1#2{#2} \def\path#1{#1}\fi

\bibitem{goldsmid10-TEintro}
H.~J. Goldsmid, Introduction to Thermoelectricity, Springer-Verlag, Berlin,
  2010.

\bibitem{vining09-TEinconvenient}
C.~B. Vining, An inconvenient truth about thermoelectrics, Nat. Mater. 8 (2009)
  83.

\bibitem{snyder08}
G.~J. Snyder, E.~S. Toberer, Complex thermoelectric materials, Nat. Mater. 7
  (2008) 105.

\bibitem{mahan96-bestTE}
G.~D. Mahan, J.~O. Sofo, The best thermoelectric, Proc. Natl. Acad. Sci. U.S.A.
  93 (1996) 7436.

\bibitem{zhao2014ultralow}
L.~D. Zhao, S.~H. Lo, Y.~Zhang, H.~Sun, G.~Tan, C.~Uher, C.~Wolverton, V.~P.
  Dravid, M.~G. Kanatzidis, Ultralow thermal conductivity and high
  thermoelectric figure of merit in \textrm{SnSe} crystals, Nature 508 (2014)
  374.

\bibitem{takahashi12}
R.~Takahashi, S.~Murakami, Thermoelectric transport in topological insulators,
  Semicond. Sci. Technol 27 (2012) 124005.

\bibitem{chen2013}
X.~Chen, D.~Parker, D.~J. Singh, Importance of non-parabolic band effects in
  the thermoelectric properties of semiconductors, Sci. Rep. 3 (2013) 3168.

\bibitem{wickramaratne15-inse}
D.~Wickramaratne, F.~Zahid, R.~K. Lake, Electronic and thermoelectric
  properties of van der \text{Waals} materials with ring-shaped valence bands,
  J. Appl. Phys. 118 (2015) 075101.

\bibitem{rudderham20}
C.~Rudderham, M.~Jesse, Analysis of simple scattering models on the
  thermoelectric performance of analytical electron dispersions, J. Appl. Phys.
  127 (2020) 065105.

\bibitem{kuroki07}
K.~Kuroki, R.~Arita, “\text{Pudding Mold}” band drives large thermopower in
  \text{Na}$_x$\text{Co}\text{O}$_2$, J. Phys. Soc. Jpn. 76 (2007) 083707.

\bibitem{kuroki13}
H.~Usui, K.~Suzuki, K.~Kuroki, S.~Nakano, K.~Kudo, M.~Nohara, Large
  \text{Seebeck} effect in electron-doped \text{FeAs}${}_{2}$ driven by a
  quasi-one-dimensional pudding-mold-type band, Phys. Rev. B 88 (2013) 075140.

\bibitem{wang14}
X.~G. Wang, L.~Wang, J.~Liu, L.~M. Peng, Camel-back band-induced power factor
  enhancement of thermoelectric lead-tellurium from \text{Boltzmann} transport
  calculations, Appl Phys. Lett. 104~(13) (2014) 132106.

\bibitem{rudderham21}
C.~Rudderham, J.~Maassen, Ab initio thermoelectric calculations of ring-shaped
  bands in two-dimensional \text{Bi}$_2$\text{Te}$_3$,
  \text{Bi}$_2$\text{Se}$_3$, and \text{Sb}$_2$\text{Te}$_3$: Comparison of
  scattering approximations, Phys. Rev. B 103 (2021) 165406.

\bibitem{mecholsky14-bandwarping}
N.~A. Mecholsky, L.~Resca, I.~L. Pegg, M.~Fornari, Theory of band warping and
  its effects on thermoelectronic transport properties, Phys. Rev. B (2014)
  155131.

\bibitem{kuroki17}
H.~Usui, K.~Kuroki, Enhanced power factor and reduced lorenz number in the
  \text{Wiedemann-Franz} law due to pudding mold type band structure, J. Appl.
  Phys. 121 (2017) 165101.

\bibitem{wolverton19}
E.~B. Isaacs, C.~Wolverton, Remarkable thermoelectric performance in
  \text{BaPdS}$_2$ via pudding-mold band structure, band convergence, and
  ultralow lattice thermal conductivity, Phys. Rev. Materials 3 (2019) 015403.

\bibitem{wei20}
S.~Wei, C.~Wang, S.~Fan, G.~Gao, Strain tunable pudding-mold-type band
  structure and thermoelectric properties of \text{SnP}$_3$ monolayer, J. Appl.
  Phys. 127 (2020) 155103.

\bibitem{mahan89-zt}
G.~Mahan, Figure of merit for thermoelectrics, J. Appl. Phys. 65 (1989) 1578.

\bibitem{sofo94-optgap}
J.~O. Sofo, G.~D. Mahan, Optimum band gap of a thermoelectric material, Phys.
  Rev. B 49 (1994) 4565.

\bibitem{hasdeo19}
E.~H. Hasdeo, L.~P.~A. Krisna, M.~Y. Hanna, B.~E. Gunara, N.~T. Hung, A.~R.~T.
  Nugraha, Optimal band gap for improved thermoelectric performance of two
  dimensional \text{Dirac} materials, J. Appl. Phys. 126 (2019) 035109.

\bibitem{chasmar1959thermoelectric}
R.~P. Chasmar, R.~Stratton, The thermoelectric figure of merit and its relation
  to thermoelectric generators, J. Electron. Control 7 (1959) 52--72.

\bibitem{hicks93-qweff}
L.~D. Hicks, M.~S. Dresselhaus, Effect of quantum-well structures on the
  thermoelectric figure of merit, Phys. Rev. B 47 (1993) 12727.

\bibitem{hicks96-TEexp}
L.~D. Hicks, T.~C. Harman, X.~Sun, M.~S. Dresselhaus, Experimental study of the
  effect of quantum-well structures on the thermoelectric figure of merit,
  Phys. Rev. B 53~(16) (1996) R10493.

\bibitem{heremans2013thermoelectrics}
J.~P. Heremans, M.~S. Dresselhaus, L.~E. Bell, D.~T. Morelli, When
  thermoelectrics reached the nanoscale, Nat. Nanotechnol. 8 (2013) 471--473.

\bibitem{hung16-quantum}
N.~T. Hung, E.~H. Hasdeo, A.~R.~T. Nugraha, M.~S. Dresselhaus, R.~Saito,
  Quantum effects in the thermoelectric power factor of low-dimensional
  semiconductors, Phys. Rev. Lett. 117 (2016) 036602.

\bibitem{QE-2017}
P.~Giannozzi, O.~Andreussi, T.~Brumme, O.~Bunau, M.~{Buongiorno Nardelli},
  M.~Calandra, R.~Car, C.~Cavazzoni, D.~Ceresoli, M.~Cococcioni, N.~Colonna,
  I.~Carnimeo, A.~{Dal Corso}, S.~{de Gironcoli}, P.~Delugas, R.~A. DiStasio,
  A.~Ferretti, A.~Floris, G.~Fratesi, G.~Fugallo, R.~Gebauer, U.~Gerstmann,
  F.~Giustino, T.~Gorni, J.~Jia, M.~Kawamura, H.-Y. Ko, A.~Kokalj,
  E.~K\"u{\c{c}}\"ukbenli, M.~Lazzeri, M.~Marsili, N.~Marzari, F.~Mauri, N.~L.
  Nguyen, H.-V. Nguyen, A.~{Otero-de-la-Roza}, L.~Paulatto, S.~Ponc{\'{e}},
  D.~Rocca, R.~Sabatini, B.~Santra, M.~Schlipf, A.~P. Seitsonen, A.~Smogunov,
  I.~Timrov, T.~Thonhauser, P.~Umari, N.~Vast, X.~Wu, S.~Baroni, Advanced
  capabilities for materials modelling with \text{Quantum ESPRESSO}, J. Phys.
  Condens. Matter 29 (2017) 465901.

\bibitem{madsen2006boltztrap}
G.~K.~H. Madsen, D.~J. Singh, \text{BoltzTraP. A} code for calculating
  band-structure dependent quantities, Comput. Phys. Commun. 175~(1) (2006)
  67--71.

\bibitem{BoltzTraP2}
G.~K.~H. Madsen, J.~Carrete, M.~J. Verstraete, {BoltzTraP2}, a program for
  interpolating band structures and calculating semi-classical transport
  coefficients, Comput. Phys. Commun. 231 (2018) 140--145.

\bibitem{githubpuddingTE}
Python codes for obtaining all the results from this paper are available at
  \url{http://github.com/jyestama/PuddingTE}.

\bibitem{ashcroft}
N.~Ashcroft, D.~Mermin, Solid State Physics, Harcourt, Orlando, 1976.

\bibitem{hung17inse}
N.~T. Hung, A.~R.~T. Nugraha, R.~Saito, Two-dimensional \textrm{InSe} as a
  potential thermoelectric material, Appl. Phys. Lett. 111~(9) (2017) 092107.

\bibitem{schiedemantel03}
T.~J. Scheidemantel, C.~Ambrosch-Draxl, T.~Thonhauser, J.~V. Badding, J.~O.
  Sofo, Transport coefficients from first-principles calculations, Phys. Rev. B
  68 (2003) 125210.

\bibitem{singh10}
D.~J. Singh, Doping-dependent thermopower of \text{PbTe} from boltzmann
  transport calculations, Phys. Rev. B 81 (2010) 195217.

\bibitem{2020SciPy-NMeth}
P.~Virtanen, R.~Gommers, T.~E. Oliphant, M.~Haberland, T.~Reddy, D.~Cournapeau,
  E.~Burovski, P.~Peterson, W.~Weckesser, J.~Bright, S.~J. {van der Walt},
  M.~Brett, J.~Wilson, K.~J. Millman, N.~Mayorov, A.~R.~J. Nelson, E.~Jones,
  R.~Kern, E.~Larson, C.~J. Carey, {\.I}.~Polat, Y.~Feng, E.~W. Moore,
  J.~{VanderPlas}, D.~Laxalde, J.~Perktold, R.~Cimrman, I.~Henriksen, E.~A.
  Quintero, C.~R. Harris, A.~M. Archibald, A.~H. Ribeiro, F.~Pedregosa, P.~{van
  Mulbregt}, {SciPy 1.0 Contributors}, {{SciPy} 1.0: Fundamental Algorithms for
  Scientific Computing in Python}, Nat. Methods 17 (2020) 261--272.

\bibitem{aflow2012}
S.~Curtarolo, W.~Setyawan, G.~L. Hart, M.~Jahnatek, R.~V. Chepulskii, R.~H.
  Taylor, S.~Wang, J.~Xue, K.~Yang, O.~Levy, M.~J. Mehl, H.~T. Stokes, D.~O.
  Demchenko, D.~Morgan, Aflow: An automatic framework for high-throughput
  materials discovery, Comput. Mater. Sci. 58 (2012) 218--226.

\bibitem{aflowlib2012}
S.~Curtarolo, W.~Setyawan, S.~Wang, J.~Xue, K.~Yang, R.~H. Taylor, L.~J.
  Nelson, G.~L. Hart, S.~Sanvito, M.~Buongiorno-Nardelli, N.~Mingo, O.~Levy,
  Aflowlib.org: A distributed materials properties repository from
  high-throughput ab initio calculations, Comput. Mater. Sci. 58 (2012)
  227--235.

\bibitem{Li2017}
J.~Li, Z.~Chen, X.~Zhang, Y.~Sun, J.~Yang, Y.~Pei, Electronic origin of the
  high thermoelectric performance of gete among the p-type group iv
  monotellurides, NPG Asia Mater. 9 (2017) e353--e353.

\bibitem{Hinterleitner2019}
B.~Hinterleitner, I.~Knapp, M.~Poneder, Y.~Shi, H.~M{\"u}ller, G.~Eguchi,
  C.~Eisenmenger-Sittner, M.~St{\"o}ger-Pollach, Y.~Kakefuda, N.~Kawamoto,
  Q.~Guo, T.~Baba, T.~Mori, S.~Ullah, X.-Q. Chen, E.~Bauer, Thermoelectric
  performance of a metastable thin-film heusler alloy, Nature 576 (2019)
  85--90.

\bibitem{Zhang2017}
J.~Zhang, L.~Song, A.~Mamakhel, M.~R.~V. J{\o}rgensen, B.~B. Iversen,
  High-performance low-cost n-type se-doped mg3sb2-based zintl compounds for
  thermoelectric application, Chem. Mater. 29 (2017) 5371--5383.

\bibitem{pei2011convergence}
Y.~Pei, X.~Shi, A.~LaLonde, H.~Wang, L.~Chen, G.~J. Snyder, Convergence of
  electronic bands for high performance bulk thermoelectrics, Nature 473 (2011)
  66--69.

\bibitem{pslibrary}
Pseudopotentials used in the first principle calculation are taken from~\url{https://dalcorso.github.io/pslibrary/}

\bibitem{dalcorso2014}
A.~Dal~Corso, Pseudopotentials periodic table: From H to Pu, Comput. Mater. Sci. 95 (2014) 337-350.

\bibitem{usui2014}
H.~Usui, K.~Kuroki, S.~Nakano, K.~Kudo, M.~Nohara, Pudding-Mold-Type Band as an Origin of the Large Seebeck Coefficient Coexisting with Metallic Conductivity in Carrier-Doped FeAs2 and PtSe2, J. Electron. Mater. 43 (2014) 10.1007/s11664-013-2823-5.   

\end{thebibliography}

\end{document}